\documentclass[twocolumn,showpacs,preprintnumbers,amsmath,amssymb]{revtex4-1}
\usepackage[dvipdfmx]{graphicx}
\usepackage{dcolumn}
\usepackage{bm}

\newcommand{\be}{\begin{eqnarray}}
\newcommand{\ee}{\end{eqnarray}}

\begin{document}


\title{Phase diagram of dipolar hard-core bosons on honeycomb lattice}

\author{Takashi Nakafuji, Takeshi Ito, Yuya Nagamori, and Ikuo Ichinose}
\affiliation{
Department of Applied Physics, Nagoya Institute of Technology, 
Nagoya, 466-8555, Japan}

\date{\today}

\begin{abstract}
In this paper, we study phase diagrams of dipolar hard-core boson gases 
on the honeycomb lattice.
The system is described by the Haldane-Bose-Hubbard model with 
complex hopping amplitudes and the nearest neighbor repulsion.
By using the slave-particle representation of the hard-core bosons and
also the path-integral quantum Monte-Carlo simulations, we 
investigate the system and to show that the systems have a rich phase diagram. 
There are Mott, superfluid, chiral superfluid, and sublattice chiral superfluid phases
as well as the density-wave phase.
We also found that there exists a coexisting phase of superfluid and chiral superfluid.
Critical behaviors of the phase transitions are also clarified.
\end{abstract}

\maketitle

\section{Introduction}

In the last few decades, the ultra-cold atomic gases are one of the most
intensively studied field in physics both experimentally and theoretically.
Advantage of the system of ultra-cold atoms is the strong-controllability
and versatility.
In particular, the ultra-cold atoms in an optical lattice (OL) paves the new way for
atomic quantum simulators to investigate the strongly-correlated systems in
the condensed matter physics, the gauge theoretical models
in the high-energy physics, etc \cite{coldatoms}.
Furthermore, academic theoretical models are realized by the atomic systems
in an OL, and possible new phenomena/state in theoretical consideration are
to be realized by the atomic systems.
In this work, we shall study one of these models, i.e., the Haldane-Bose-Hubbard
model (HBHM) in the honeycomb lattice.

The Haldane mode of fermions in the honeycomb lattice was introduced 
as a model that has similar
states at certain fillings to the state of the quantum Hall effect (QHE) \cite{Haldane}. 
It exhibits non-trivial topological properties as a result of the complex 
hopping parameters.
Bosonic counterpart of the Haldane model was introduced, 
and its ground-state phase and low-energy excitations were studied at 
unit filling \cite{BHM1}. 
In the previous work, we studied the HBHM in the honeycomb lattice by means 
of the extended Monte-Carlo (MC) simulations, and showed that the system
has a very rich phase diagram \cite{Kuno}.
In particular, we are interested in the effect of the complex next-nearest-neighbor (NNN)
hopping amplitude, which is now experimentally feasible by the recent advance
in generating an artificial magnetic flux in the OL \cite{mg_optical,Honeycomb_ex}.
In addition to the ordinary superfluid (SF), some exotic SF, which is called chiral SF
(CSF), forms for sufficiently large NNN hoppings, whereas the bosonic state of the QHE does not form.
This result confirmed the previous works on the HBHM \cite{BHM1}.
Furthermore, we found that there appears a co-existing phase of the SF and CSF
if the pattern of the complex NNN hoppings is changed to that of
the model called the modified Haldane model \cite{Kuno}.
In the CSF and SF+CSF phases, topological excitations, vortices, play an important role.
We also studied the HBHM in the cylinder geometry and found that  some exotic
state forms near the boundary.

In this paper, we shall study a system of dipolar hard-core boson in the 
honeycomb lattice.
Effect of the dipole is expressed by the strong NN repulsion \cite{DDI}.
We show that interplay of the NN repulsion and the complex NNN hopping generate a
very rich phase diagram. 
This result partially stems from the quantum uncertainty relation between 
the particle number and the phase of the boson operator.

This paper is organized as follows.
In Sec. II, we introduce the hard-core HBHM and the slave-particle representation
for the hard-core boson.
For the complex NNN hopping, we consider two cases, one is that of the original 
Hubbard model and the other is called the modified Hubbard model.
By integrating out the quantum fluctuations of the densities, we derive an effective
low-energy model that is bounded from below.
In Sec. III, by the MC simulation of the effective model, we study the phase diagram
in detail.
There are various phases in the phase diagram, and we clarify the physical
properties of the phases and critical behavior of the phase transitions.
Section IV is devoted for discussion and conclusion.

\section{Models in slave-particle representation and numerical methods}

\begin{figure}[t]
\centering
\includegraphics[width=7cm]{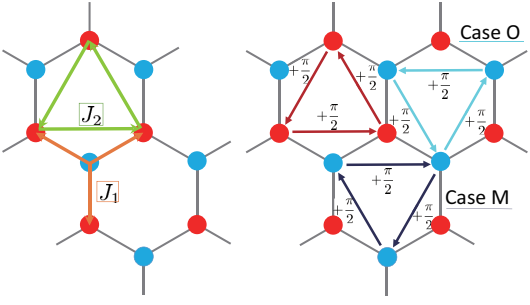}
\caption{(Color online) Hopping parameters of the HBHM in Eq.(\ref{HBHM}).
The phase $\phi={\pi \over 2}$ in the direction of arrows.
Please notice that the directions of arrows are different in the Case ${\cal O}$
and Case ${\cal M}$.
}
\label{model}
\end{figure}

Hamiltonian of the HBHM is given as follows,
\begin{eqnarray}
H_{\rm HBH} &=& -J_1\sum_{\langle i,j \rangle}(a^\dagger_ia_j+a^\dagger_ja_i)
-J_2\sum_{\langle\langle i,j \rangle\rangle}e^{i\phi}
(a^\dagger_ia_j+a^\dagger_ja_i) \nonumber \\
&&+U\sum_in_i^2+V\sum_{\langle i,j \rangle}n_in_j
-\mu\sum_in_i,
\label{HBHM}
\end{eqnarray}
where $a_i \ (a^\dagger_i)$ is the annihilation (creation) operator at site $i$,
$n_i=a^\dagger_ia_i$, $J_1$ and $J_2$ are the nearest-neighbor (NN) and 
the next-NN (NNN) hopping amplitudes, respectively.
$U(>0)$ and $V(>0)$ represent on-site and NN repulsions, respectively, 
$\phi$ is a phase of the NNN hopping amplitude, which will be specified shortly,
and $\mu$ is the chemical potential. 
The NN repulsion $V$ comes from the interaction between dipoles of atoms
whose orientations are perpendicular to the honeycomb lattice. 
In the present numerical study, we mostly employ the grand-canonical ensemble.
The bosons in the system $H_{\rm HBH}$, Eq.(\ref{HBHM}), have hard-core nature.
The hopping parameters are depicted in Fig.\ref{model}.
For the phase $\phi$, we consider two different cases as shown in Fig.\ref{model}.
The Case ${\cal O}$ corresponds to the original HBHM, whereas 
the Case ${\cal M}$ is sometimes called a modified HBHM \cite{modified}.
These two models have a different phase structure for certain parameter
regions as we show in this work.

We consider the large-$U$ case in this paper.
In order to treat the hard-core boson, we employ the slave-particle
representation.
The creation operator $a_i$ is represented as follows in terms of the 
hole operator $h_i$ and the particle operator $b_i$,
\begin{equation}
a_i=h^\dagger_ib_i,
\label{HCB}
\end{equation}
with the constraint
\begin{equation}
(b^\dagger_ib_i+h^\dagger_ih_i)|\mbox{Phy}\rangle=|\mbox{Phy}\rangle,
\label{const}
\end{equation}
where $|\mbox{Phy}\rangle$ denotes the physical subspace of the slave particles
corresponding to the hard-core boson Hilbert space.
From Eqs.(\ref{HCB}) and (\ref{const}),
it is not difficult to show that the operator $a_i$ and $a^\dagger_i$ on the same site
satisfy the fermionic anti-commutation relation as $\{a_i, a^\dagger_i\}=1$, 
whereas the usual bosonic commutation relations as $[a_i,a^\dagger_j]=0$, etc.,
for $i\neq j$.

The system is studied by the path-integral methods developed in Ref.\cite{EMC}.
To this end, we introduce the coherent states for the `slave particles' as
\begin{eqnarray}
&& b_i|z_{bi}\rangle=z_{bi}|z_{bi}\rangle=\sqrt{\rho_{bi}} 
\ e^{i\theta_{bi}}|z_{bi}\rangle, \nonumber \\
&& h_i|z_{hi}\rangle=z_{hi}|z_{hi}\rangle=\sqrt{\rho_{hi}} \ e^{i\theta_{hi}}|z_{hi}\rangle,
\label{coherent}
\end{eqnarray}
where $\rho_{bi} \ (\rho_{hi})$ is the density of particle (hole) at site $i$,
and $\theta_{bi} \ (\theta_{hi})$ is the phase \cite{note0}.
In the path-integral calculation, we divide $\rho_{bi} \ (\rho_{hi})$ into 
its average $\bar{\rho}_{bi} \ (\bar{\rho}_{hi})$ and its quantum 
fluctuation $\delta{\rho}_{bi} \ (\delta{\rho}_{hi})$ around it, i.e.,
$\rho_{bi}=\bar{\rho}_{bi}+\delta\rho_{bi}, \ 
\rho_{hi}=\bar{\rho}_{hi}+\delta\rho_{hi}$.
The averages $\bar{\rho}_{bi}$ and $\bar{\rho}_{hi}$ are determined by the
condition that linear terms of $\delta{\rho}_{bi}$ and $\delta{\rho}_{hi}$ 
are absent in the action.
We impose the local constraint Eq.(\ref{const}), $\bar{\rho}_{bi}+\bar{\rho}_{hi}=1$.
Numerical calculation in later sections exhibits that $\bar{\rho}_{bi}$ and $\bar{\rho}_{hi}$
are actually slow variables compared to the quantum variables $\theta_{bi}$ and
$\theta_{hi}$.
Integration over $\delta\rho_{bi}$ and $\delta\rho_{hi}$ can be performed
as in the previous works without any difficulty, and a suitable
effective action, $S[\theta_b, \theta_h;\bar{\rho}_b, \bar{\rho}_h]$,
is derived.
For the MC simulation, we introduce a lattice in the imaginary-time $\tau$-direction
with the lattice spacing $\Delta\tau$ and parameterize $\bar{\rho}_{bi}$ and
$\bar{\rho}_{hi}$ as $\bar{\rho}_{bi,\ell}=\sin^2(X_{i,\ell}), \
\bar{\rho}_{hi,\ell}=\cos^2(X_{i,\ell})$, where $X_{i,\ell}$'s are angle variables and 
$(i,\ell)$ denotes site in the stacked honeycomb lattice $(\ell:\mbox{imaginary time})$.
Furthermore from Eq.(\ref{HCB}), it is obvious that the phase of $h_i$ (or $b_i$)
is redundant and therefore we put $\theta_{hi}=0$ in the practical calculation.
The partition function $Z_{\rm HBH}$ is given as follows\cite{note1},
\begin{eqnarray}
Z_{\rm HBH}=\int\prod^{N_\tau-1}_{\ell=0}\prod_i [dX_{i,\ell}d\theta_{bi,\ell}]
e^{-S}, 
\label{Z1} 
\end{eqnarray}
\begin{eqnarray}
&&S=\sum^{N_\tau-1}_{\ell=0}\Big[\sum_i-{1\over 2U_\tau\Delta \tau}
\cos(\theta_{bi,\ell+1})-\theta_{bi,\ell}) \nonumber \\
&&-{1 \over 2}J_1\Delta\tau\sum_{\langle i,j\rangle}\sin(2X_{i,\ell})
\sin(2X_{j,\ell})\cos(\theta_{bi,\ell}-\theta_{bj,\ell}) \nonumber \\
&&-{1 \over 2}J_2\Delta\tau\sum_{\langle\langle i,j\rangle\rangle}\sin(2X_{i,\ell})
\sin(2X_{j,\ell})\cos(\theta_{bi,\ell}-\theta_{bj,\ell}+\phi) \nonumber \\
&&+U'\Delta\tau\sum_i\sin^4(X_{i,\ell})
+V\Delta\tau\sum_{\langle i,j\rangle}\sin^2(X_{i,\ell})\sin^2(X_{j,\ell}) \nonumber \\
&&-\mu\sum_i\sin^2(X_{i,\ell})-\sum_i\log(\sin(2X_{i,\ell}))\Big],
\label{partitionF}
\end{eqnarray}
where $N_\tau$ is the linear size of the $\tau$-direction and is related to 
the temperature ($T$) as $N_\tau\Delta\tau=1/(k_{\rm B}T)$, and all variables are 
periodic in the $\tau$-direction.
$U_\tau$ is the parameter that controls quantum fluctuations of the phases
$\{\theta_{bi,\ell}\}$ and is related to the repulsions that control
the density fluctuations.
In the later sections, we shall consider two typical cases of $U_\tau$, i.e.,
$U_\tau=0.1$ and 10 to see the effect of the quantum fluctuations.
We will see that different phase diagrams appear depending on the value of $U_\tau$.
As the system is the hard-core gases, fairly large quantum fluctuations of the phases 
$\{\theta_{bi,\ell}\}$ are expected.
However, a certain experimental manipulation, e.g., using Rabi coupling with a 
stable reference SF of the gases, may stabilize the phase fluctuations. 
Therefore, to study the system with small $U_\tau$ is not meaningless.
The $U'$-term was introduced in order to suppress the multi-particles states
that appear as a result of the use of the coherent-state path integral.
(See the footnote \cite{note0}.)
We studied the case of $U'=10$ and $0.1$ and found that the $U'$-term
does not substantially influence the numerical results.
The last term in Eq.(\ref{partitionF}) comes from the change of variables
from $(\bar{\rho}_{bi,\ell}, \ \bar{\rho}_{hi,\ell})$ to $X_{i,\ell}$.
As the action $S$ in Eq.(\ref{partitionF})
is real and bounded from below, there exist no difficulties in performing MC simulations.
In the following sections, we shall show the numerical results and discuss 
the physical meaning of them.

\section{Results of MC simulations}

\subsection{Physical observables}

In the previous section, we derived the effective model for the HBH models of
the hard-core bosons.
In this section, we shall show the numerical results obtained by the path-integral
MC simulations.
In particular, we clarify the phase diagrams of these models.
To this end, we calculate various quantities.
Phase boundaries are identified by calculating the ``internal energy" $E$ and 
the ``specific heat" $C$ defined as,
\begin{eqnarray}
E={\langle S \rangle \over N_s}, \;\;\;
C={\langle (S-\langle S\rangle)^2 \rangle \over N_s},
\label{UC}
\end{eqnarray}
where $N_s$ is the total number of sites in the stacked honeycomb lattice
and we employ the periodic boundary condition.
We divide the honeycomb lattice into sublattices $A$ and $B$.
Average particle number in the $A$-sublattice, $\rho_A$, and  that in the $B$-sublattice,
$\rho_B$, are defined as 
\begin{equation}
\rho_A=\sum_{i\in A, \ell}\langle \sin^2(X_{i,\ell}) \rangle/N_A,  \;\;
\rho_B=\sum_{i\in B, \ell}\langle \sin^2(X_{i,\ell}) \rangle/N_B, 
\end{equation}
where $N_{A(B)}$ is the total number of sites in the stacked $A(B)$-sublattice.
Fluctuations of the particle number at each sublattice, $\Delta\rho_{A(B)}$ is 
defined similarly.

\begin{figure}[t]
\centering
\includegraphics[width=5cm]{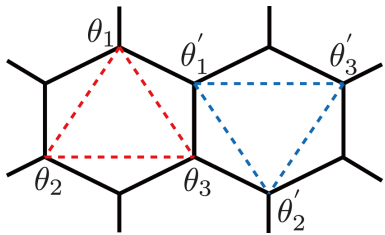}
\caption{(Color online) Parameters for the definition of vortex Eq.(\ref{VAVB}).
}
\label{vortextheta}
\end{figure}

Vortices play an important role to discuss physical properties of the states.
For the Case ${\cal O}$, the vorticity of a triangle in each sublattice is defined as follows,
\begin{eqnarray}
&& {\cal V}_A={1\over 3}\Big[\sin(\theta_3-\theta_1+{\pi \over 2})
+\sin(\theta_2-\theta_3+{\pi \over 2})  \nonumber \\
&&\hspace{1cm}+\sin(\theta_1-\theta_2+{\pi \over 2})\Big],  \nonumber \\
&& {\cal V}_B={1 \over 3}\Big[\sin(\theta'_3-\theta'_1-{\pi \over 2})
+\sin(\theta'_2-\theta'_3-{\pi \over 2})  \nonumber \\
&&\hspace{1cm}+\sin(\theta'_1-\theta'_2-{\pi \over 2})\Big], 
\label{VAVB}
\end{eqnarray}
where $\theta_m(m=1,2,3)$ and $\theta'_m(m=1,2,3)$ are shown in Fig.\ref{vortextheta}.
For the Case ${\cal M}$, $-{\pi \over 2}$ is replaced with ${\pi \over 2}$
in the definition ${\cal V}_B$ in Eq.(\ref{VAVB}).
For the 120$^o$ configurations in the sub-lattice, ${\cal V}_{A(B)}$
takes the value $\mp 0.5$.

\begin{figure}[t]
\centering
\includegraphics[width=5cm]{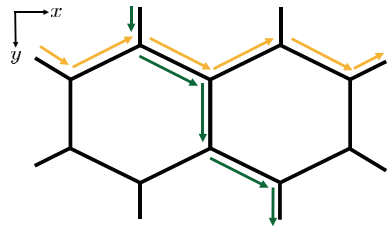}
\caption{(Color online) $x$ and $(x+y)$-directions in the honeycomb lattice.
System size in the $x$-direction ($y$-direction) is denoted by $n_x$ $(n_y)$.
}
\label{corrdirection}
\end{figure}

We also calculate correlation functions on the honeycomb lattice,
\begin{eqnarray}
&&G_{\rm x}(r)={1 \over N_s}\sum_i\langle \cos(\theta_{i+r}-\theta_i)\rangle, 
\; \mbox{in x-direction}, \label{GreenF}  \\
&&G_{\rm x+y}(r)={1 \over N_s}\sum_i\langle \cos(\theta_{i+r}-\theta_i)\rangle, 
\; \mbox{in (x+y)-direction}, \nonumber
\end{eqnarray}
where for the meaning of the x, y and x+y directions, see Fig.\ref{corrdirection}.
Among the correlations in Eq.(\ref{GreenF}), the following the NN and
NNN correlations were used in the previous works to identify phases \cite{Kuno},
\begin{eqnarray}
&&L_{\rm NN}={1 \over 3N_s}\sum_{\langle i, j\rangle}\cos (\theta_i-\theta_j),
\nonumber \\
&&C_{\rm NN}={1 \over 3N_s}\sum_{\langle i, j\rangle}\sin (\theta_i-\theta_j),
\label{LCNN}
\end{eqnarray}
where $\langle i, j\rangle$ stands for the NN sites.
Similarly for the NNN sites 
$\langle\langle i,j \rangle\rangle\in A(B)$ on the A(B)-sublattice,
\begin{eqnarray}
&&L_{\rm NNNA(B)}={1 \over 3N_s}\sum_{\langle\langle i, j\rangle\rangle\in A(B)}
\cos (\theta_i-\theta_j+{\pi \over 2}),
\nonumber \\
&&C_{\rm NNNA(B)}={1 \over 3N_s}\sum_{\langle\langle i, j\rangle\rangle\in A(B)}
\sin (\theta_i-\theta_j\pm{\pi \over 2}).
\label{LCNNN}
\end{eqnarray}
The averages of the A and B-sublattice quantities are defined naturally
\begin{eqnarray}
&&L_{\rm NNN}={1 \over 2}(L_{\rm NNNA}+L_{\rm NNNB}), \nonumber \\
&&C_{\rm NNN}={1 \over 2}(C_{\rm NNNA}+C_{\rm NNNB}).
\end{eqnarray}
Behaviors of the above quantities distinguish the SF and CSF phases.
In the SF, $\bar{\rho}=\rho_A=\rho_B$ and $L_{\rm NN}>0$.
The mean-field theory also predicts that 
$C_{\rm NNNA}=C_{\rm NNNB}=2\bar{\rho}J_2>0$, 
and $C_{\rm NN}=0$ for the SF.
On the other hand for the CSF, $C_{\rm NNNA}=C_{\rm NNNB}=-2\bar{\rho}J_2<0$
and $C_{\rm NN}\neq 0$.
More complicated states are identified by means of the above correlations.
See later discussion.

\subsection{$J_2=0$ case}

\begin{figure}[t]
\centering
\includegraphics[width=5cm]{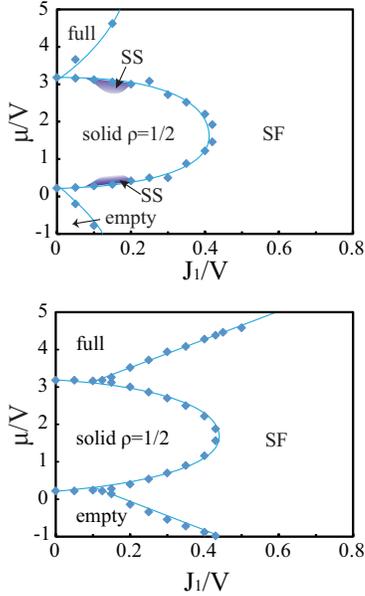}
\caption{(Color online) Phase diagram of the dipolar hard-core boson on the
honeycomb lattice with vanishing NNN hopping $J_2=0$
and $U_\tau=0.1 (10.0)$ for the upper (lower) panel.
There are four phases, empty, full, $\rho={1 \over 2}$ half-filled solid,
and superfluid (SF).
For the case $U_\tau=0.1$, supersolid (SS) forms in small but finite parameter regions.
Dots indicate the phase transition points observed by the MC simulations.
Error bars are roughly the same size with the dots.
}
\label{J20}
\end{figure}
\begin{figure}[t]
\centering
\includegraphics[width=5.4cm]{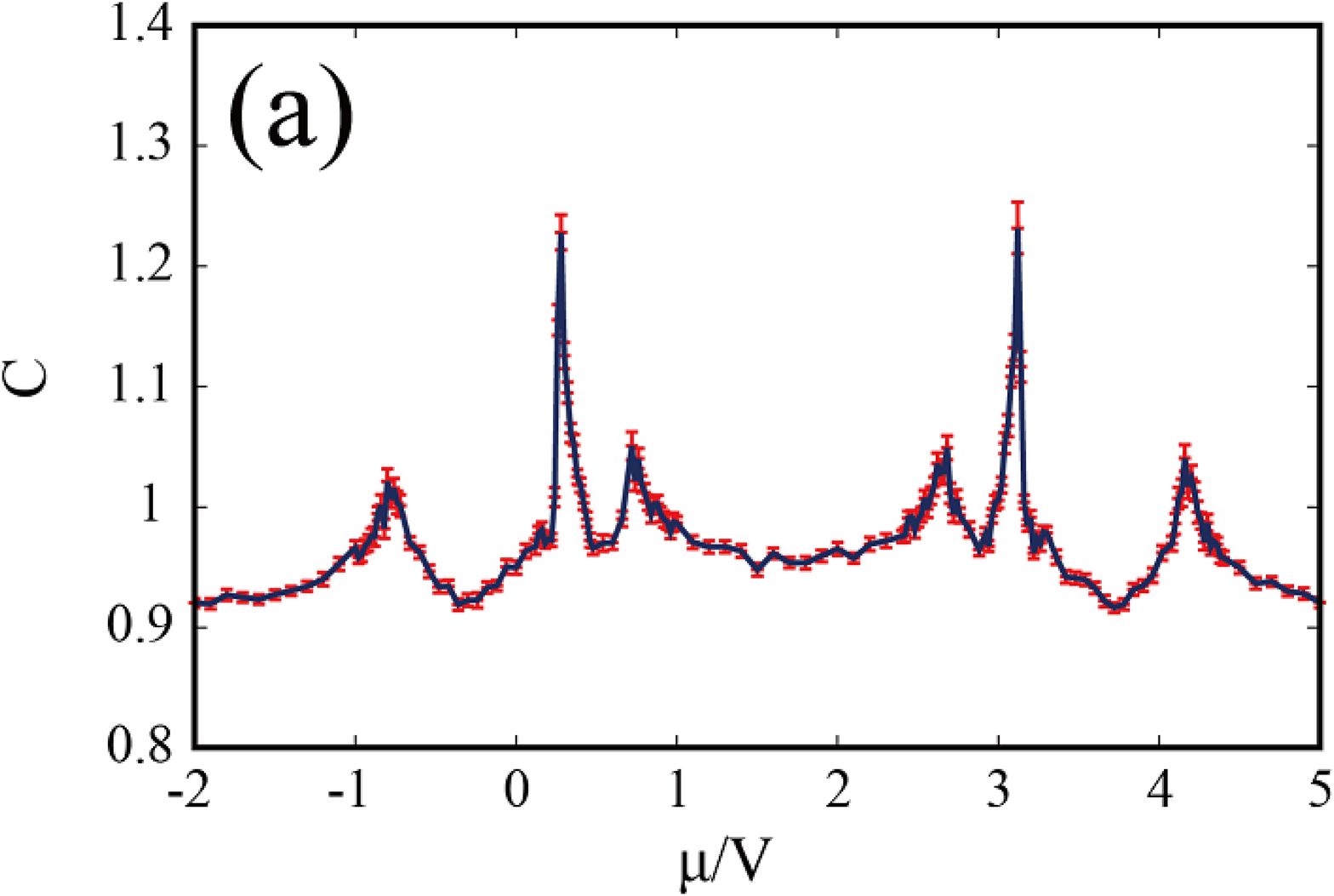}
\includegraphics[width=5cm]{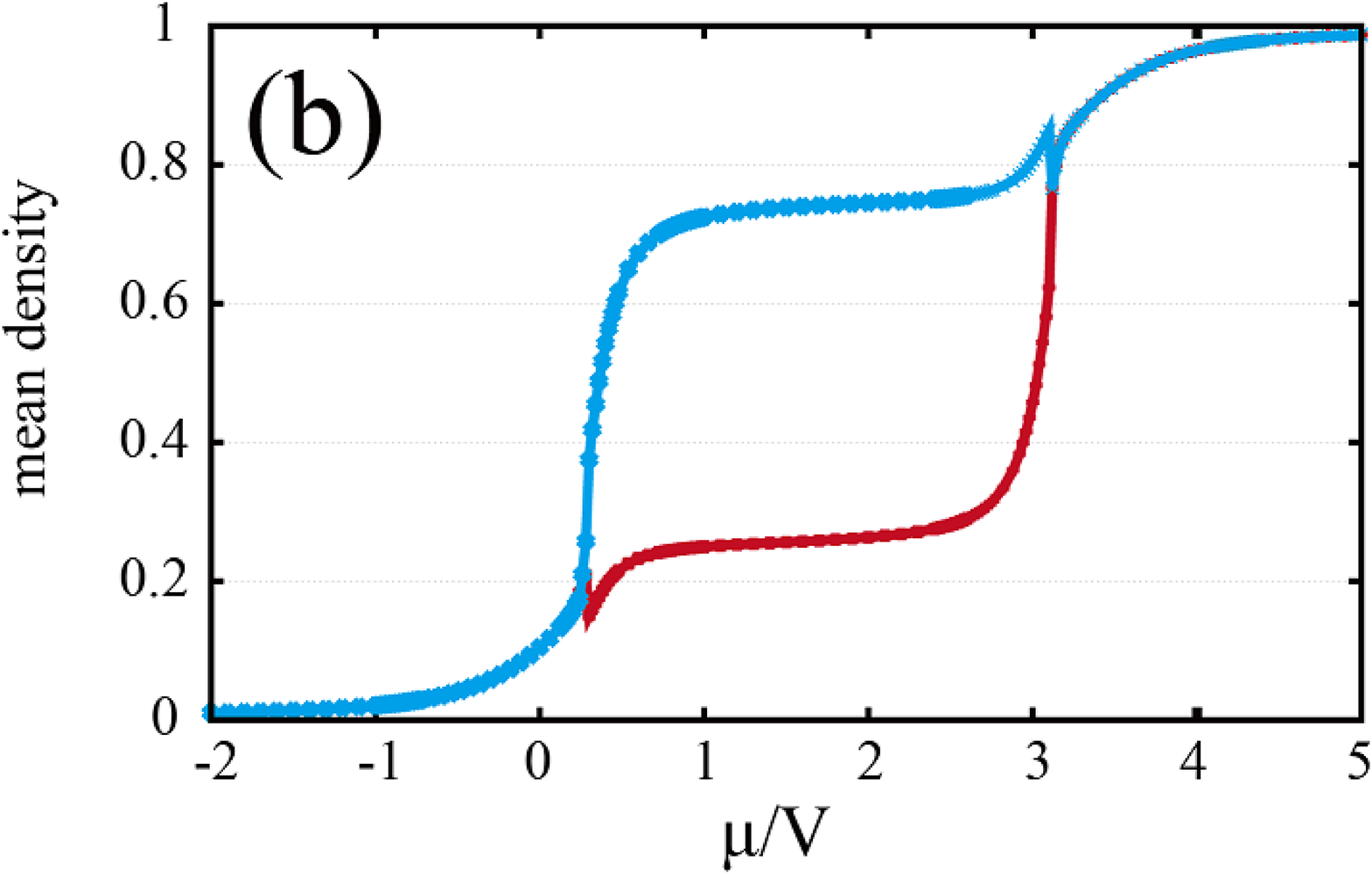}
\includegraphics[width=5cm]{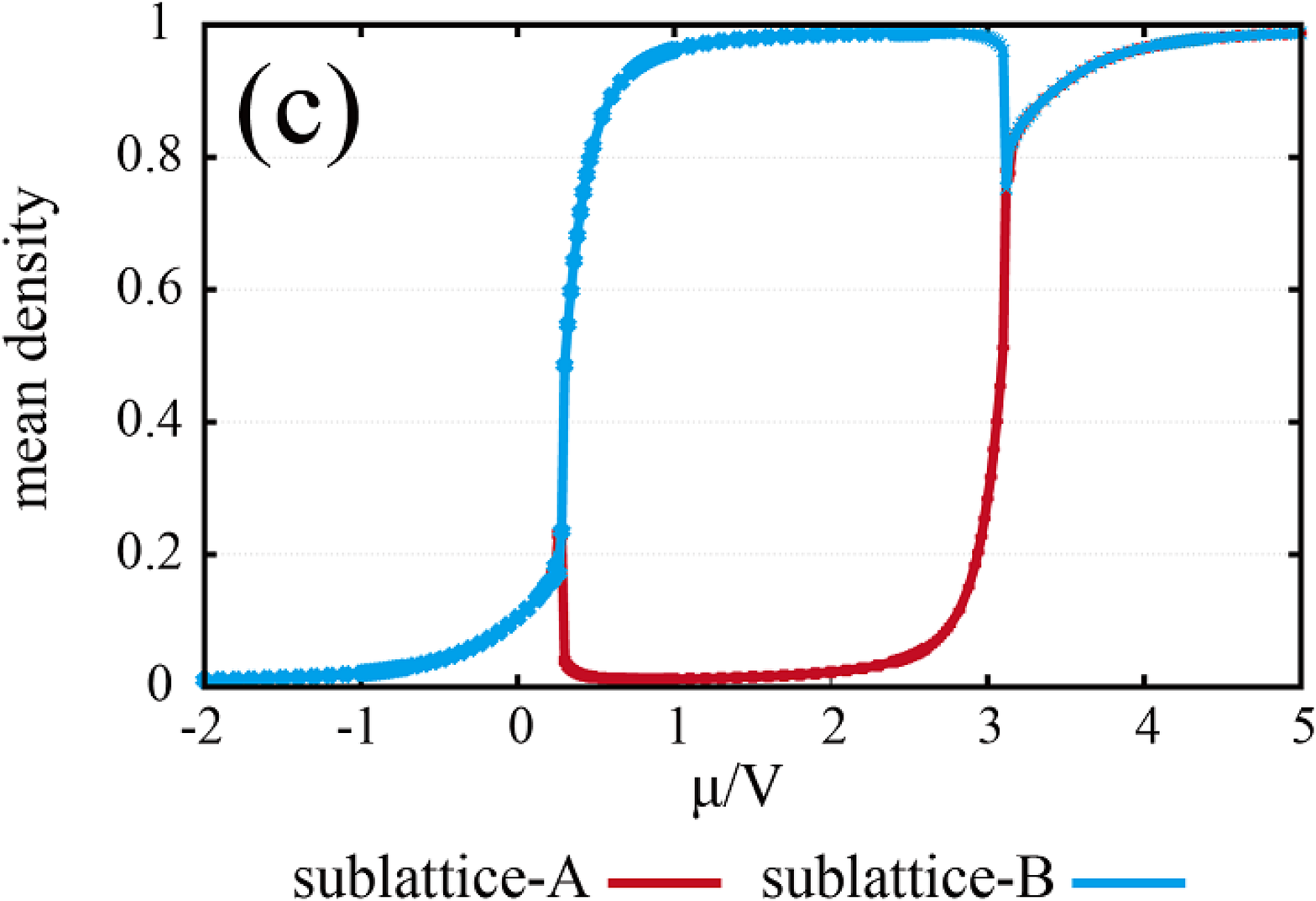}
\includegraphics[width=6.1cm]{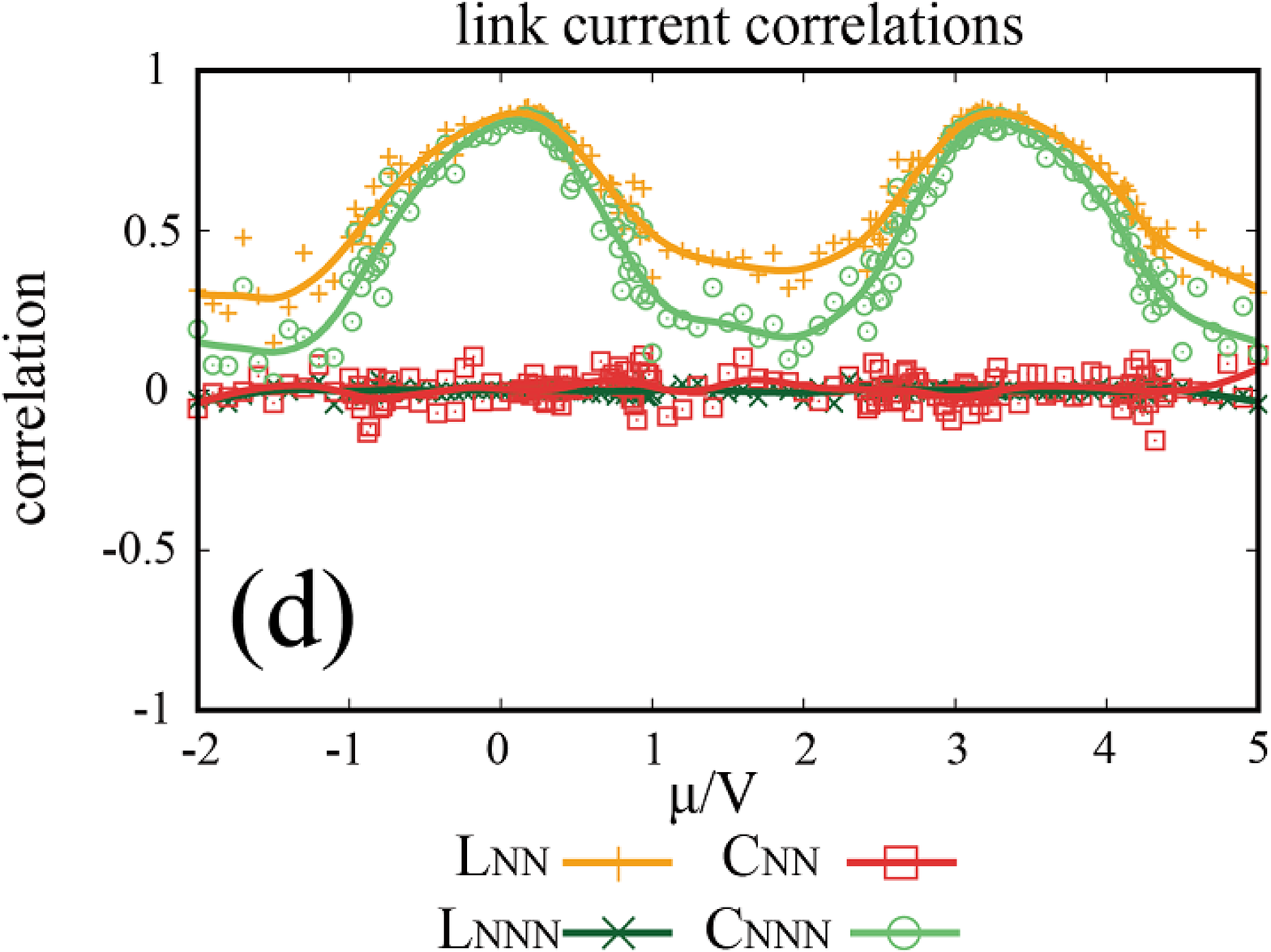}
\caption{(Color online) 
a) Specific heat $C$ for $J_1/V=0.1, \ U_\tau=0.1$ and $J_2=0$.
There are four peaks indicating the genuine phase transitions.
Two peaks at $\mu/V\simeq 0.7$ and $2.7$ seem to indicate
the existence of the supersolid (SS), since there exists a finite superfluidity
as shown in Fig.\ref{corrJ20}.
System size $n_x(=n_y)=36$.
b) The total average particle densities of the $A$ and $B$-sublattices.
In the phase of $\rho=1/2$ solid, there exists an imbalance.
c) The average particle densities of the $A$ and $B$-sublattices in single
layer in the $\tau$-direction.
d) Order parameters for $J_2=0, \ J_1/V=0.1$.
In the SF, $L_{\rm NN}$ and $C_{\rm NNN}$ have positive values.
System size $n_x=12$.
}
\label{CJ20}
\end{figure}

In the practical calculations, we put $\Delta \tau=1$, and also $V=50.0$.
As the temperature of the system $T$ is related to $\Delta\tau$ as 
$k_{\rm B}T=1/(N_\tau\Delta\tau)$, to put $\Delta \tau=1$ means
setting the energy unit to $(N_\tau k_{\rm B}T)$.
Then we are considering ultra-cold atomic systems as $V/k_{\rm B}T \sim 400$ 
for $N_\tau\sim 8$.
Estimation of the temperature $T$ will be given for a typical atomic gas
in Sec.IV.

We first study the simplest case with the vanishing NNN hopping, $J_2=0$.
To obtain the phase diagram in the $({J_1 \over V}-{\mu\over V})$-plane,
we calculate the specific heat $C$ as a function of $\mu/V$ for fixed values
of $J_1$.
Phase boundaries are determined by the locations of peaks in $C$.
The obtained phase diagrams are shown in Fig.\ref{J20} for $U_\tau=0.1$ and $10$.
In both cases, there are four phases, empty, full, half-filled Mott insulator (MI),
and superfluid (SF).
The half-filled MI is nothing but the sublattice density wave (DW).
The calculated specific heat $C$ is shown in Fig.\ref{CJ20}-a).
The sharp peaks in $C$ exhibit that error bars of the locations of the phase boundaries 
in Fig.\ref{J20} is very small.
In Fig.\ref{CJ20}-b) and c), the average particle densities in the $A$ and 
$B$-sublattices are shown.
In the half-filled solid state, only $A$ or $B$-sublattice is occupied and the 
other is empty.
In Fig.\ref{CJ20}-d), we show the calculations of the various correlations. 
In the SF, $L_{\rm NN}$ and $C_{\rm NNN}$ have a positive value.
On the other hand, $C_{\rm NN}$ and $L_{\rm NNN}$ vanish.
This result comes from the coherent Bose condensation of the boson at
the vanishing momentum.
In the $\rho=1/2$-solid, the $L_{\rm NN}$ and $C_{\rm NNN}$ have a small but 
finite value because of the short-range correlation.

The phase diagram in Fig.\ref{J20} should be compared to
the previously obtained phase diagram by the quantum MC simulation 
using the stochastic series expansion \cite{SSE}.
In fact, the phase diagram in Fig.\ref{J20}, in particular that of the case $U_\tau=10$, is
in good agreement with that in Ref.\cite{SSE}.
There exist three phase boundaries all of which coincide quantitatively
in the two phase diagrams.
This result indicates the reliability of the present method. 

\begin{figure}[t]
\centering
\includegraphics[width=5cm]{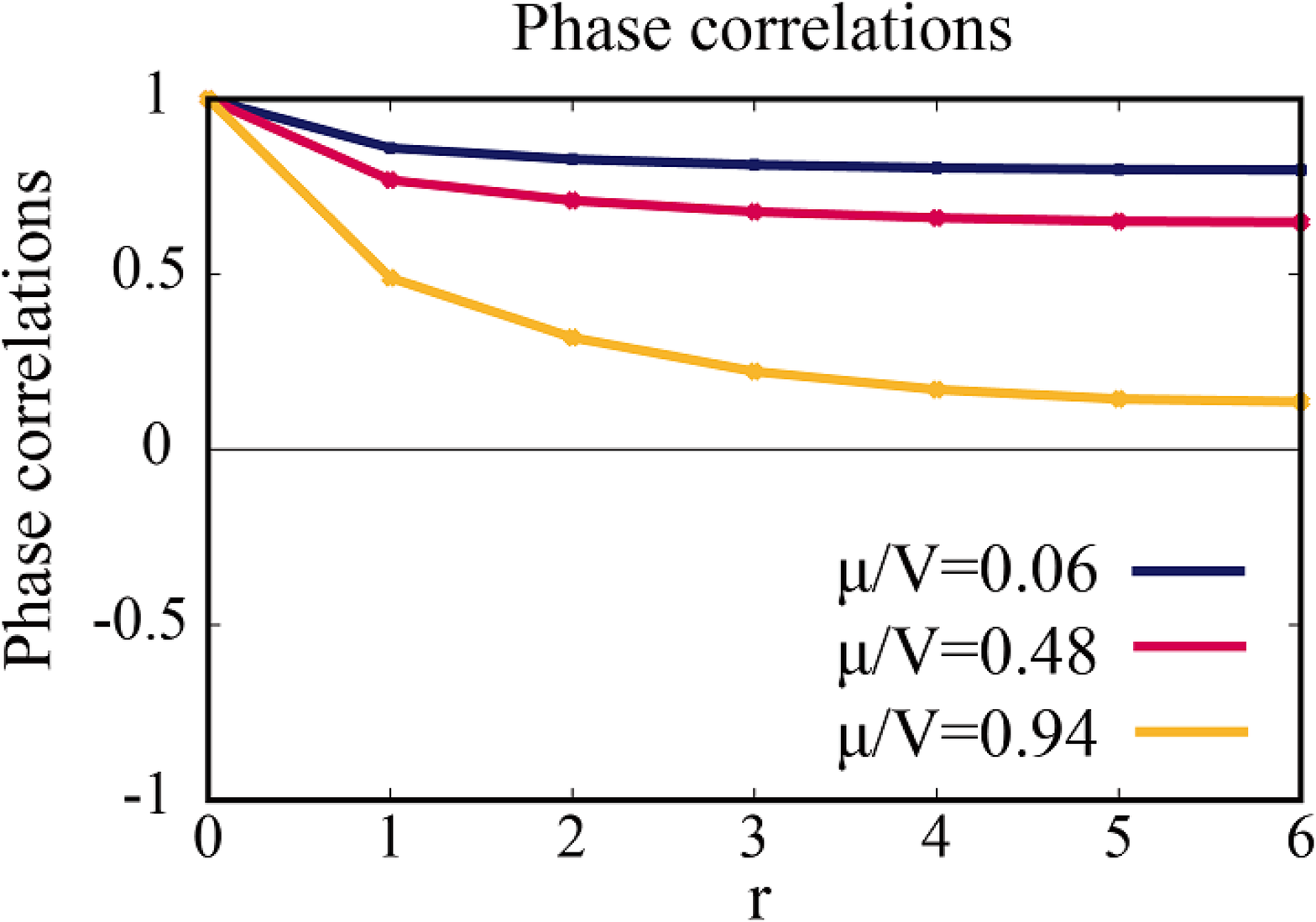}
\includegraphics[width=3.5cm]{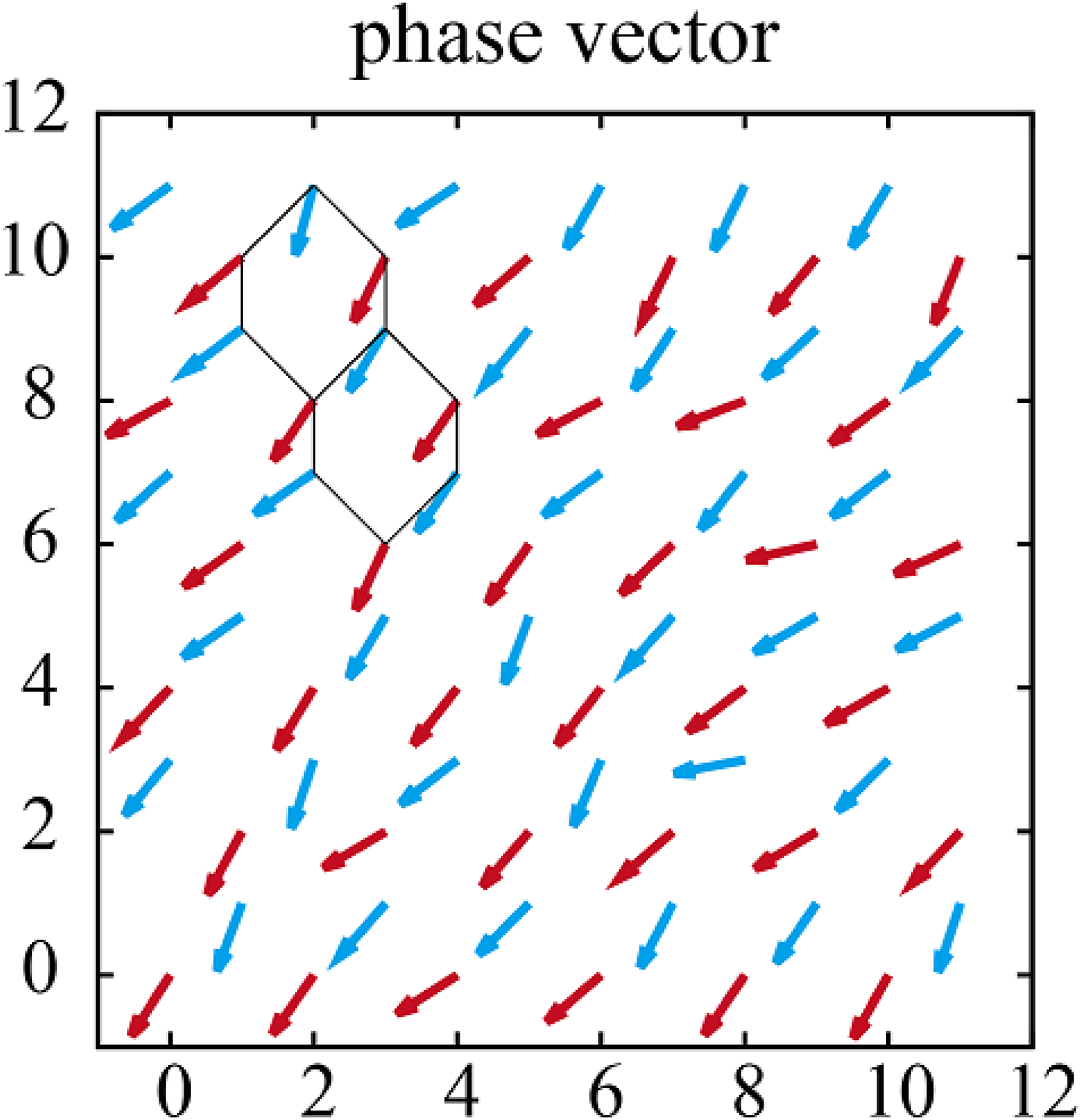}
\caption{(Color online) (Left panel) Correlation function of the SF in the $x$-direction
for the case of vanishing NNN hopping $J_2=0, J_1/V=0.1$ and $U_\tau=0.1$. 
(Right panel) Phase of the Bose condensation at each site of the honeycomb lattice.
$J_1/V=0.28, \ \mu/V=3$.
System size $n_x=12$.
}
\label{corrJ20}
\end{figure}

In Fig.\ref{corrJ20}, the correlation function $G_{\rm x}(r)$ that measures 
the SF density, and snapshots of the phases $\{\theta_i\}$ are also shown.
It is obvious that the long-range order of the phase degrees of freedom of the
Bose condensate exists.
The appearance of the solid with $\rho=1/2$ reveals that the NNN repulsion $V$
is strong enough compared with the NN one with $U$, and as a result, one 
of the sublattices is filled with particles and the other is totally empty.
In the SF phase, on the other hand, the average particle densities 
on the A and B-sublattices are the same as the hopping term dominates
the repulsions.
The parameter region of the SF depends on the value of $U_\tau$.
For smaller $U_\tau$, the correlation of $\{\theta_i\}$ in the $\tau$-direction
increases, and therefore the SF is enlarged.

The calculated specific heat $C$ in Fig.\ref{CJ20} shows that there are two 
small peaks at $\mu/V\simeq 0.7$ and $2.7$.
The correlation function $G_{\rm x}(r)$ in Fig.\ref{corrJ20}
show that there exists a finite SF correlation at $J_1/V=0.1$ and $\mu/V=0.48$,
which is located inside the $\rho=1/2$ solid.
This result indicates the existence of the supersolid (SS) \cite{SS}.
Density profile in Fig.\ref{CJ20} indicates that the A-sublattice filling
is fractional there whereas the B-sublattice is totally empty.
From this observation, we conclude that the SS forms as result of the `hole doping' 
(or `particle doping')to the $\rho=1/2$-solid.
Similar phenomenon was observed, e.g., the system of the hard-core boson in 
the triangular lattice \cite{ozawa}.

We investigated other values of $J_1$ up to the system size $n_x=18$ and found that 
the signal of the SS is finite, $L_{\rm NN}>0$, but is getting weak.
Clear phase boundary of the SS is rather difficult to obtain from the measure of 
the specific heat $C$ as only a moderate small peak exists in the relevant parameter
region even for large system size.
See the phase diagram in Fig.\ref{J20}.

\begin{figure}[t]
\centering
\includegraphics[width=8cm]{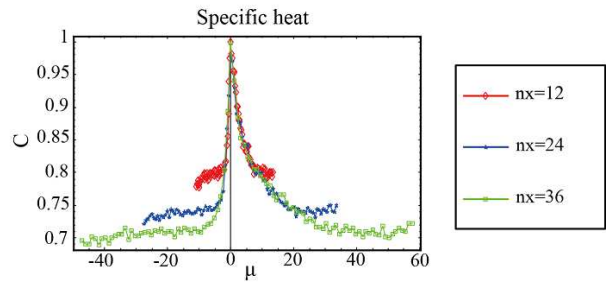}
\caption{(Color online) Scaling function $\Phi(x)$ of specific heat FSS for 
the phase transition between the $\rho=1/2$ state and the SF. $J_2=0$. 
The critical exponents are $\nu=0.75$, $\sigma=0.11$ and $\mu_\infty/V=0.27$.
The result indicates that the phase transition is of second order.
}
\label{CsFSS}
\end{figure}

Order of phase transitions is verified by the finite-size scaling (FSS) hypothesis 
of the specific heat \cite{FSS}. 
For a second-order phase transition, the calculated specific heat 
of the system linear size $L$, $C_L(\epsilon)$, satisfies the following FSS law,
\begin{eqnarray}
C_L(\epsilon)&=&L^{\sigma/\nu}\Phi(L^{1/\nu}\epsilon^\nu), \nonumber \\
\epsilon&=&(g-g_\infty)/g_\infty
\label{FSS1}
\end{eqnarray}
where $\Phi(x)$ is a scaling function, $g$ generally denotes coupling constant  
and $g_\infty$ is the critical coupling for the system size $L\rightarrow \infty$.
$\nu$ and $\sigma$ are critical exponent that characterize the phase transition.
In the phase transition between the $\rho=1/2$ solid and the SF, the `coupling constant'
$g=\mu$.
We calculated $C_L$ for $L(=n_x)=12, 24$ and 36 and applied the FSS hypothesis
for the $\rho=1/2$-SF phase transition located at 
$(J_1/V\simeq 0.1, \mu/V\simeq 0.2)$.
The result shown in Fig.\ref{CsFSS} shows that the FSS hypothesis is satisfied 
quite well with the critical exponents $\nu=0.98, \ \sigma=0.064$ and 
$\mu_\infty/V=0.27$.
Other phase transitions in the phase diagram of Fig.\ref{J20} are of second order.


\subsection{$J_2>0$ cases}

\begin{figure}[h]
\centering
\includegraphics[width=6.5cm]{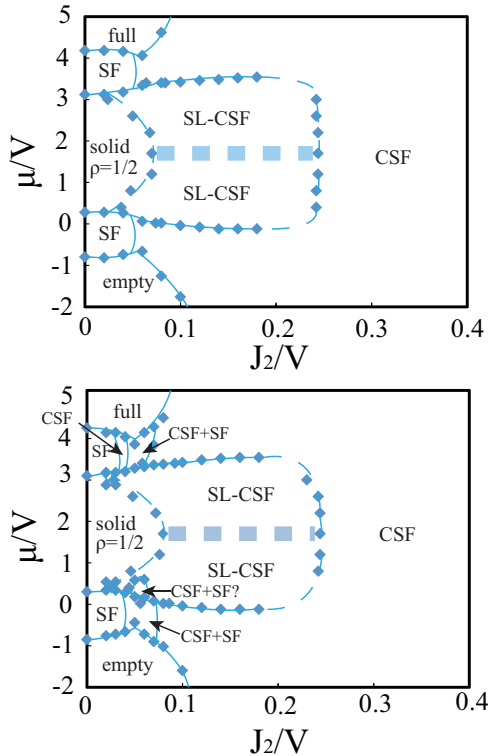}
\caption{(Color online) Phase diagram for $U_\tau=0.1$, $J_1/V=0.1$ and $V=50$.
The model of the Case ${\cal O}$ (upper panel) has empty, full, SF, 
$\rho={1 \over 2}$ solid, CSF and sublattice (SL)-CSF phases.
On the other hand, the Case ${\cal M}$ (lower panel) has a complex phase diagram, and  
there exists an additional coexistence phase of the SF and CSF.
In both cases, the transition between two SL-CSFs exhibits only a broad peak in $C$,
whereas there exists clear difference in the vortex configuration as shown in
Fig.\ref{vortexSLJ=5J2=3}.
Error bars are roughly the same size with the dots.
}
\label{PDJ1=5}
\end{figure}

\begin{figure}[h]
\centering
\includegraphics[width=4.2cm]{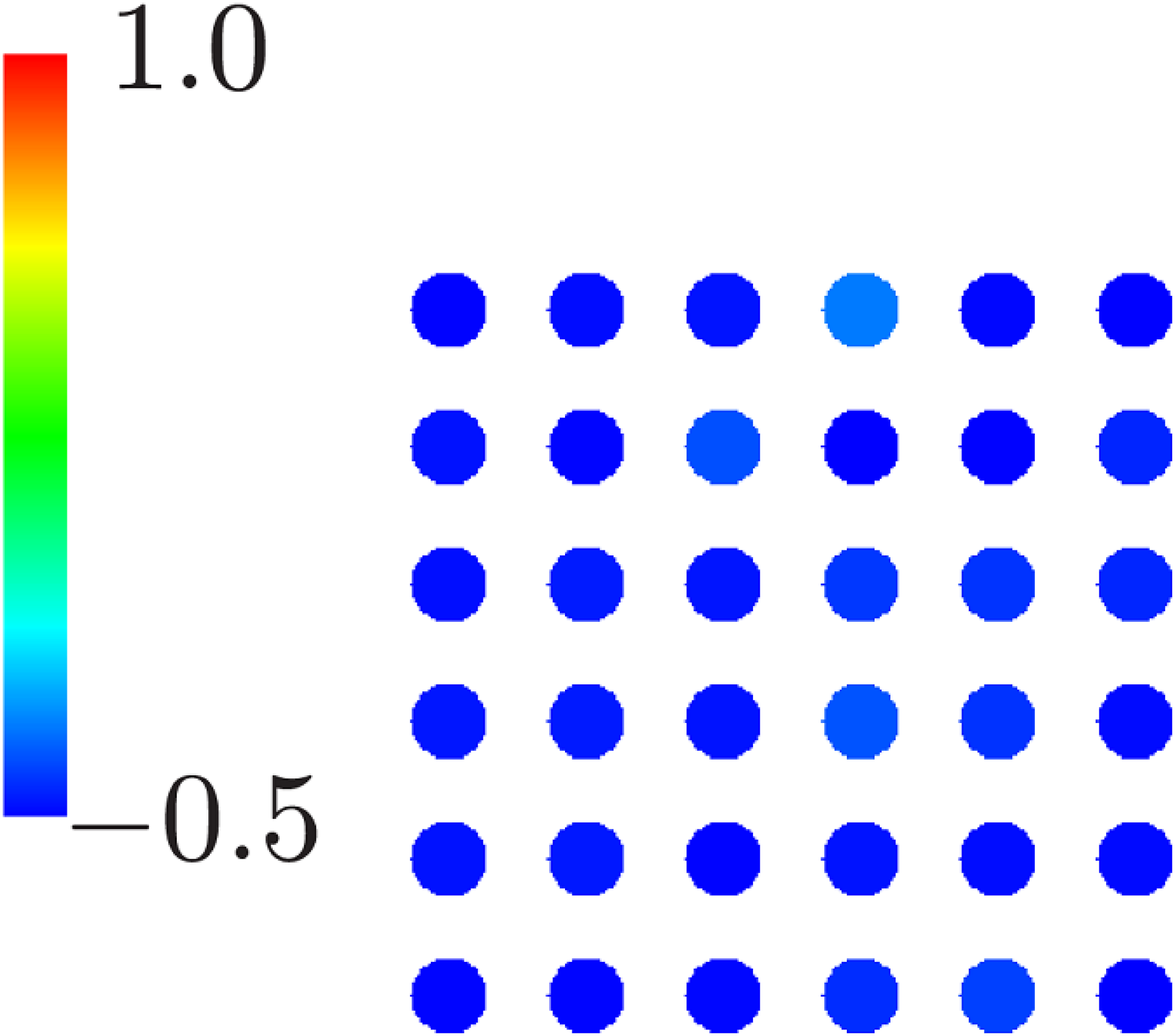} 
\includegraphics[width=4.2cm]{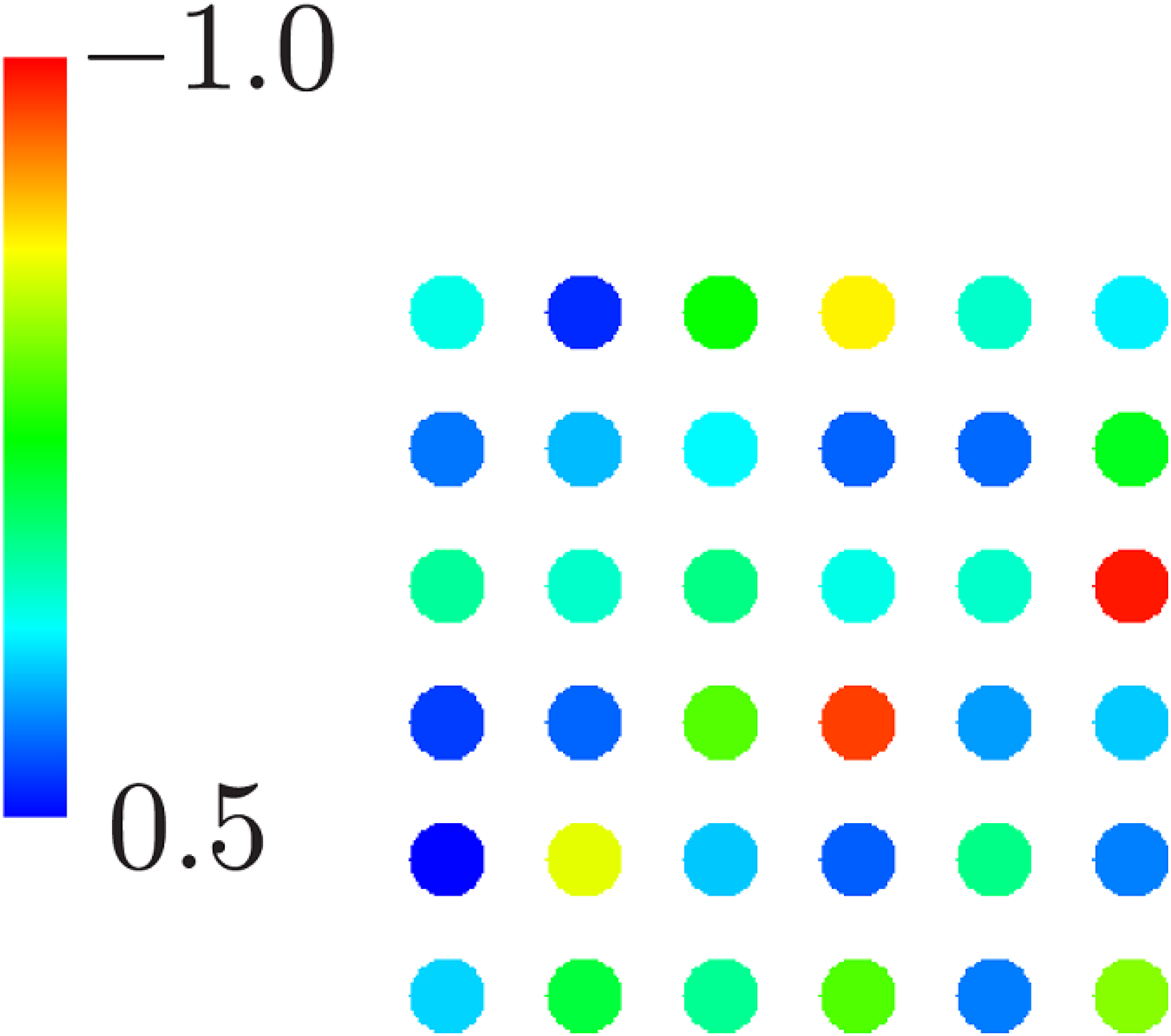}
\caption{(Color online) Vortex in A and B-sublattices in the SL-CSF.
$J_1/V=0.1, \ J_2/V=0.06$ and $\mu/V=0.48$.
In the A-sublattice (left panel), 120$^o$ configuration forms but in the B-sublattice (right panel),
no specific order of the phase exists.
In the other SL-CSF in the phase diagram in Fig.\ref{PDJ1=5}, in
the B-sublattice, -120$^o$ configuration forms but in the A-sublattice,
no specific order.
System size $n_x=12$.
}
\label{vortexSLJ=5J2=3}
\end{figure}

\begin{figure}[h]
\centering
\includegraphics[width=6cm]{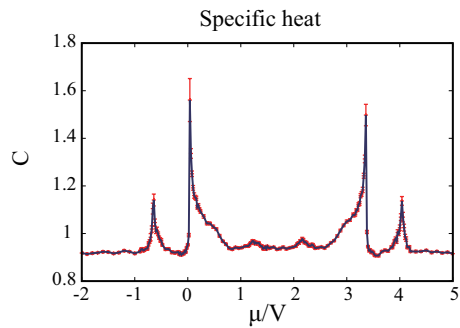}
\caption{(Color online) Specific heat $C$ for Case ${\cal O}$ with $J_1/V=0.1$ 
and $J_2/V=0.06$.
There are four clear peaks in $C$ corresponding to the four phase transition
shown in Fig.\ref{PDJ1=5}, i.e., 
``empty $\rightarrow$ CSF $\rightarrow$ SL-CSF $\rightarrow$ $\rho=1/2$ 
$\rightarrow$ SL-CSF' $\rightarrow$ CSF  $\rightarrow$ full"
Small peaks at $\mu/V \simeq 1.1$ and $2.1$ are reminiscence of the SS
existing in the $\rho=1/2$ state for the case $J_2=0$.
System size $n_x=36$.
}
\label{COJ2=3}
\end{figure}

\begin{figure}[h]
\centering
\includegraphics[width=8cm]{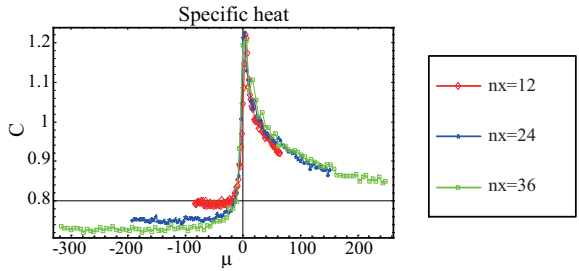}
\caption{(Color online) Scaling function of specific heat FSS for the phase transition 
between the CSF and SL-SF. 
Case ${\cal O}$ with $J_1/V=0.1$ and $J_2/V=0.06$. 
The estimated critical exponents are $\nu=0.82$, $\sigma=0.080$, and 
$\mu_\infty/V\simeq 0.036$.
The result indicates that the phase transition is of second order.
}
\label{CsFSS2}
\end{figure}
\begin{figure}[h]
\centering
\includegraphics[width=5cm]{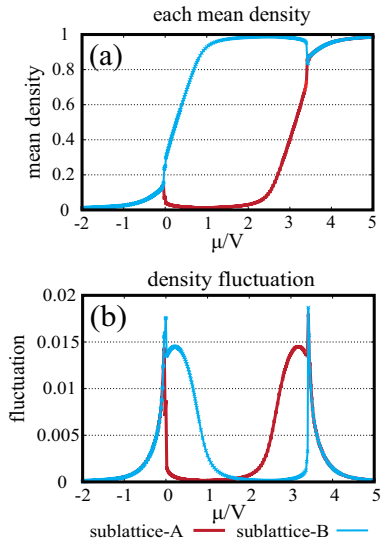}
\caption{(Color online) a) Density of the sublattice particles.
b) Density fluctuations in the A and B-sublattices,
and various order parameters for the Case ${\cal O}$ and 
$U_\tau=0.1, \ J_1/V=0.1, \ J_2/V=0.1$.
In the SL-CSF, the particle number of the chiral condensate fluctuates to form 
the phase coherence.
Atoms in the other sublattice are almost in the empty or full state without 
density fluctuations. System size $n_x=12$.
}
\label{resultJ1=5}
\end{figure}
\begin{figure}[h]
\centering
\includegraphics[width=5.5cm]{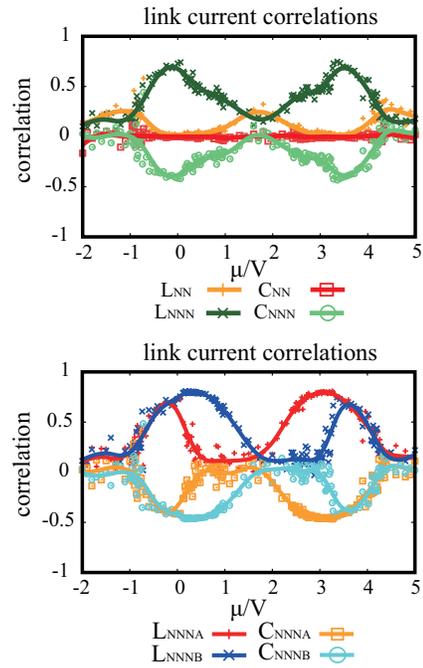}
\caption{(Color online) 
(Upper panel) Total link and current correlations in Case ${\cal O}$. 
$U_\tau=0.1, \ J_1/V=0.1, \ J_2/V=0.06$.
(Lower panel) Sublattice link and current correlations in Case ${\cal O}$
with the above parameters. 
From the phase diagram in Fig.\ref{PDJ1=5}, the phase transition takes place
at $\mu/V \sim -1, 0, 1.5$ and $3.2$, i.e.,
``empty $\rightarrow$ CSF $\rightarrow$ SL-CSF $\rightarrow$
SL-CSF' $\rightarrow$ CSF".
See Fig.\ref{PDJ1=5} (upper panel).
In the CSF, $C_{\rm NNN}<0$. System size $n_x=12$.
}
\label{currJ1=5}
\end{figure}
\begin{figure}[h]
\centering
\includegraphics[width=4.7cm]{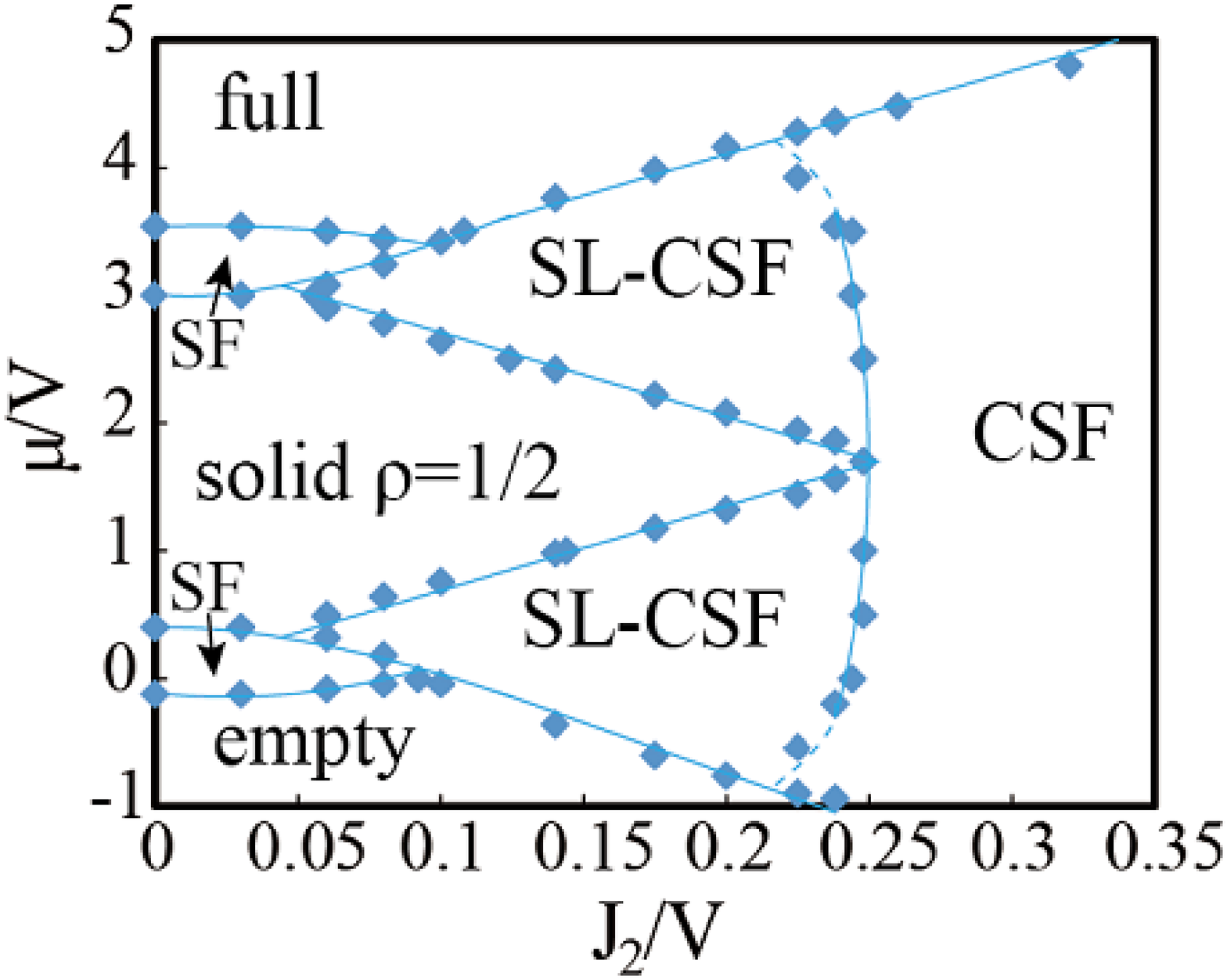} \\
\hspace{-1cm}\includegraphics[width=5.6cm]{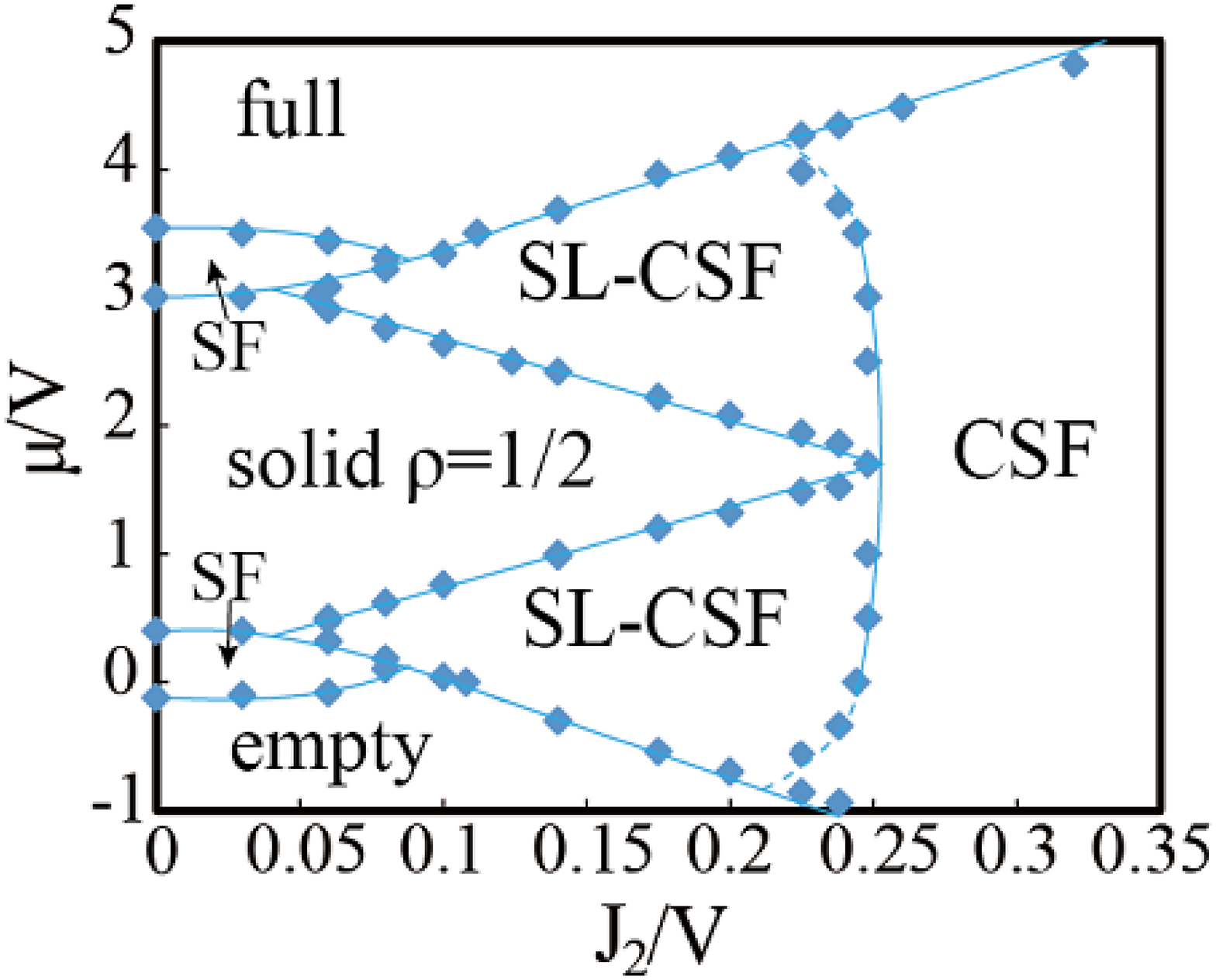}
\caption{(Color online) Phase diagram for $U_\tau=10$, $J_1/V=0.2$ and $V=50$.
The model of the Case ${\cal O}$ (upper panel) 
and the Case ${\cal M}$ (lower panel) have similar phase diagrams.
}
\label{PDJ1=5_2}
\end{figure}

In this section, we shall study the phase diagram of the Case ${\cal O}$ and also 
Case ${\cal M}$ with $J_2>0$.
First we fix the NN hopping parameter as $J_1/V=0.1$ and 
obtain the phase diagram in the $(\mu/V$-$J_2/V)$-plane by the MC simulation.
Phase boundaries are determined by calculating the specific heat $C$,
and phases are identified by calculating various physical order parameters.
The obtained phase diagrams for $U_\tau=0.1$ and $J_1/V=0.1$ 
are shown in Fig.\ref{PDJ1=5}.
In the phase diagrams of the Case ${\cal O}$, there are two phases in addition to 
the empty, full, SF and $\rho=1/2$ solid states, i.e.,  the chiral SF state (CSF)
and the sublattice (SL)-CSF.
In the CSF, the phases $\{\theta_i\}$ in both the A and B-sublattices 
form the $120^o$ configurations, whereas the SL-CSF, 
ether A or B-sublattice has the $120^o$-order and the other sublattice
has a small filling and no phase coherence.
See the vortex configuration shown in Fig.\ref{vortexSLJ=5J2=3}.
In the A-sublattice, the 120$^o$ configuration forms, whereas
in the B-sublattice, there exists no phase coherence.
Formation of the SL-CSF obviously stems from the NNN repulsion, although
increase of the NNN hopping amplitude prefers the CSF.

In the practical calculation, 
we measured the specific heat $C$ by increasing (or decreasing) the chemical potential 
$\mu$ with fixed $J_2$ or by increasing (or decreasing) $J_2$ with fixed $\mu$ 
as some of the phase boundaries are almost parallel to the constant $\mu/J_2$ line.
Typical behavior of the specific heat is shown in Fig.\ref{COJ2=3} for the Case ${\cal O}$.
The calculated specific heat $C$ for $J_2/V=0.06$ in Fig.\ref{COJ2=3} shows that
there are six peaks at $\mu/V\simeq -0.6, \ 0.0, \ 1.2, \ 2.2, \ 3.2$ and 4.0.
As the phase diagram in Fig.\ref{PDJ1=5} indicates, the phase transitions are
``empty $\rightarrow$ CSF $\rightarrow$ SL-CSF $\rightarrow$ $\rho=1/2$ 
$\rightarrow$ SL-CSF' $\rightarrow$ CSF  $\rightarrow$ full".
By the FSS hypothesis for the specific heat $C_L(\epsilon)$,
the critical exponents are to be estimated if the phase transition is of second order.
See Fig.\ref{CsFSS2} for the scaling function $\Phi(x)$ of the CSF-SL-CSF
phase transition at $J_1/V=0.1, \ J_2/V=0.06$ and $\mu/V\simeq 0$.
The result indicates that this phase transition is of second order. 

The average particle densities on the A and B-sublattices, and various order parameters are shown in Figs.\ref{resultJ1=5} and \ref{currJ1=5}, 
from which each phase in the phase diagram was identified.
The calculations of Fig.\ref{currJ1=5} indicate that there exist four phase boundaries
at $\mu/V \sim -1, 0, 1.5$ and $3.2$.
In the CSF, $C_{\rm NNNA}<0$ and $C_{\rm NNNB}<0$, whereas in 
the A-sublattice CSF, $C_{\rm NNNA}<0$ and $C_{\rm NNNB}\sim 0$,
and similarly for the B-sublattice CSF.
From the above observation, the phase boundaries are
``empty $\rightarrow$ CSF $\rightarrow$ SL-CSF $\rightarrow$
SL-CSF' $\rightarrow$ CSF".
See the phase diagram in Fig.\ref{PDJ1=5} (upper panel).
Finally,
the above calculations clearly show that for the SL-CSF with the coherent phase
to form, the fluctuations
in the particle number on that sublattice is needed as the uncertainty relation
between the number and phase suggests.

\begin{figure}[h]
\centering
\includegraphics[width=4.5cm]{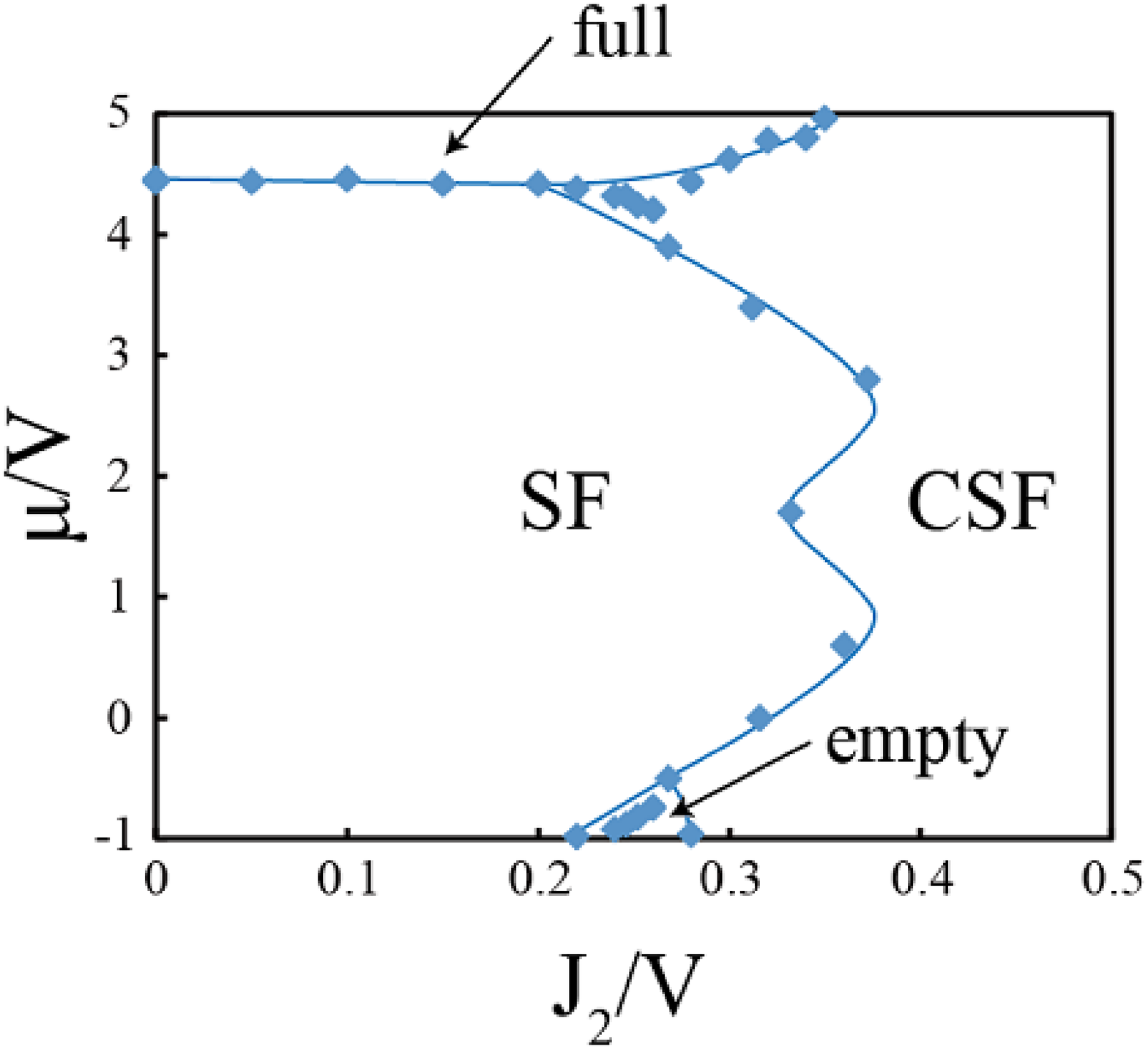}
\includegraphics[width=4.5cm]{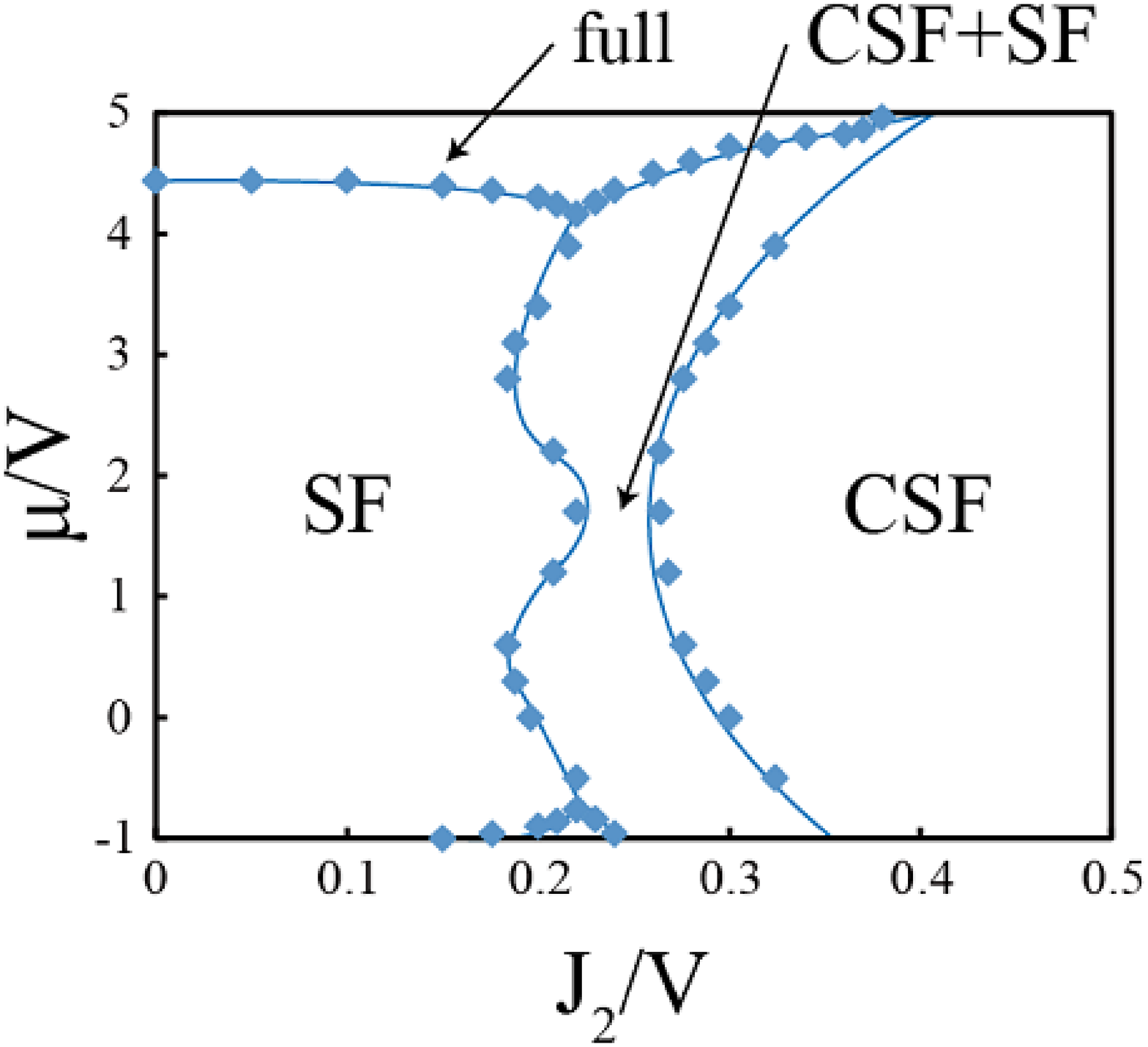}
\caption{(Color online) Phase diagram for $U_\tau=10$, $J_1/V=0.45$ and $V=50$.
Upper panel is the Case ${\cal O}$ and lower is the Case ${\cal M}$.
}
\label{PDJ1=22.5}
\end{figure}
\begin{figure}[h]
\centering
\includegraphics[width=5cm]{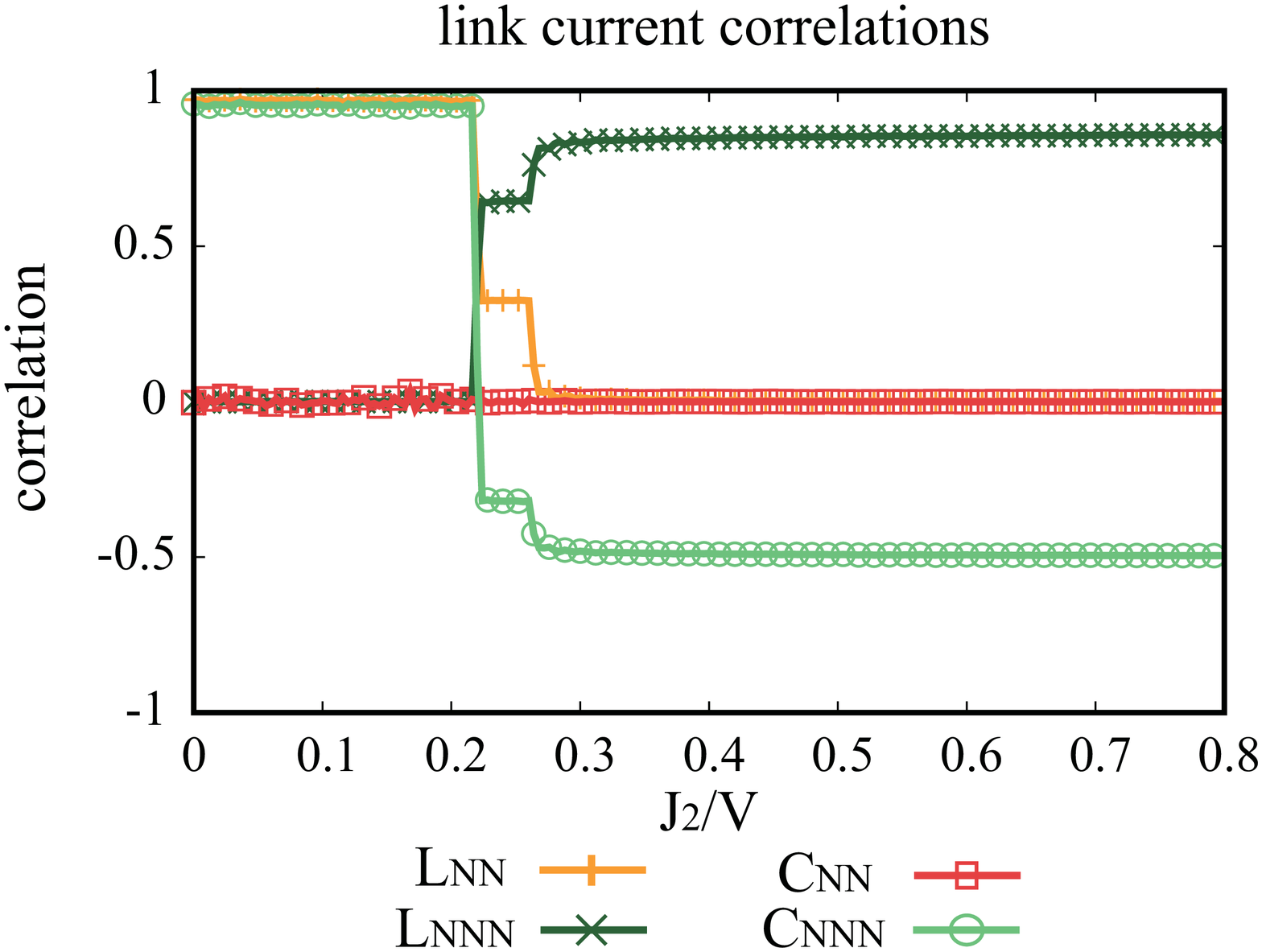}
\includegraphics[width=5cm]{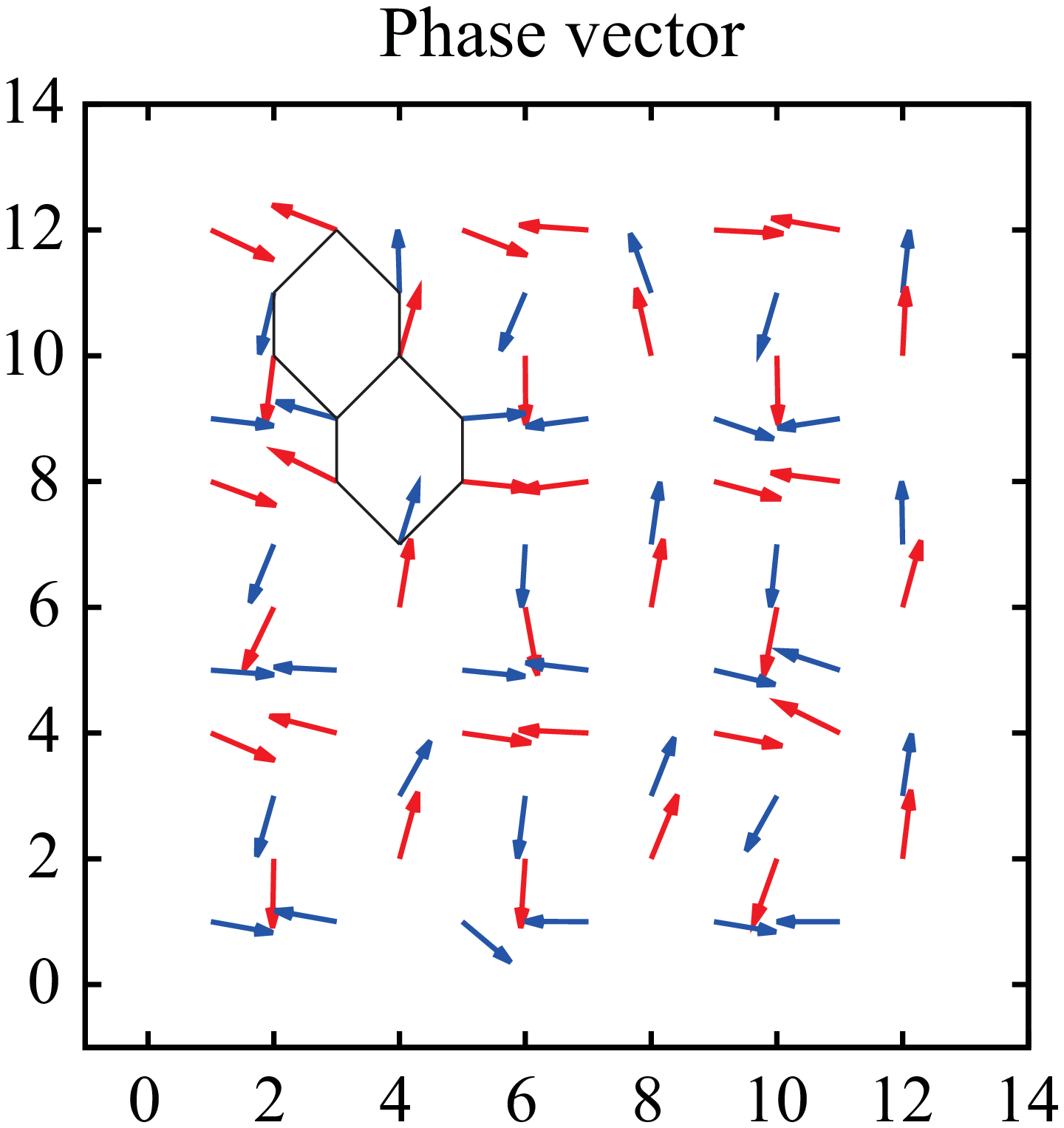}
\includegraphics[width=5cm]{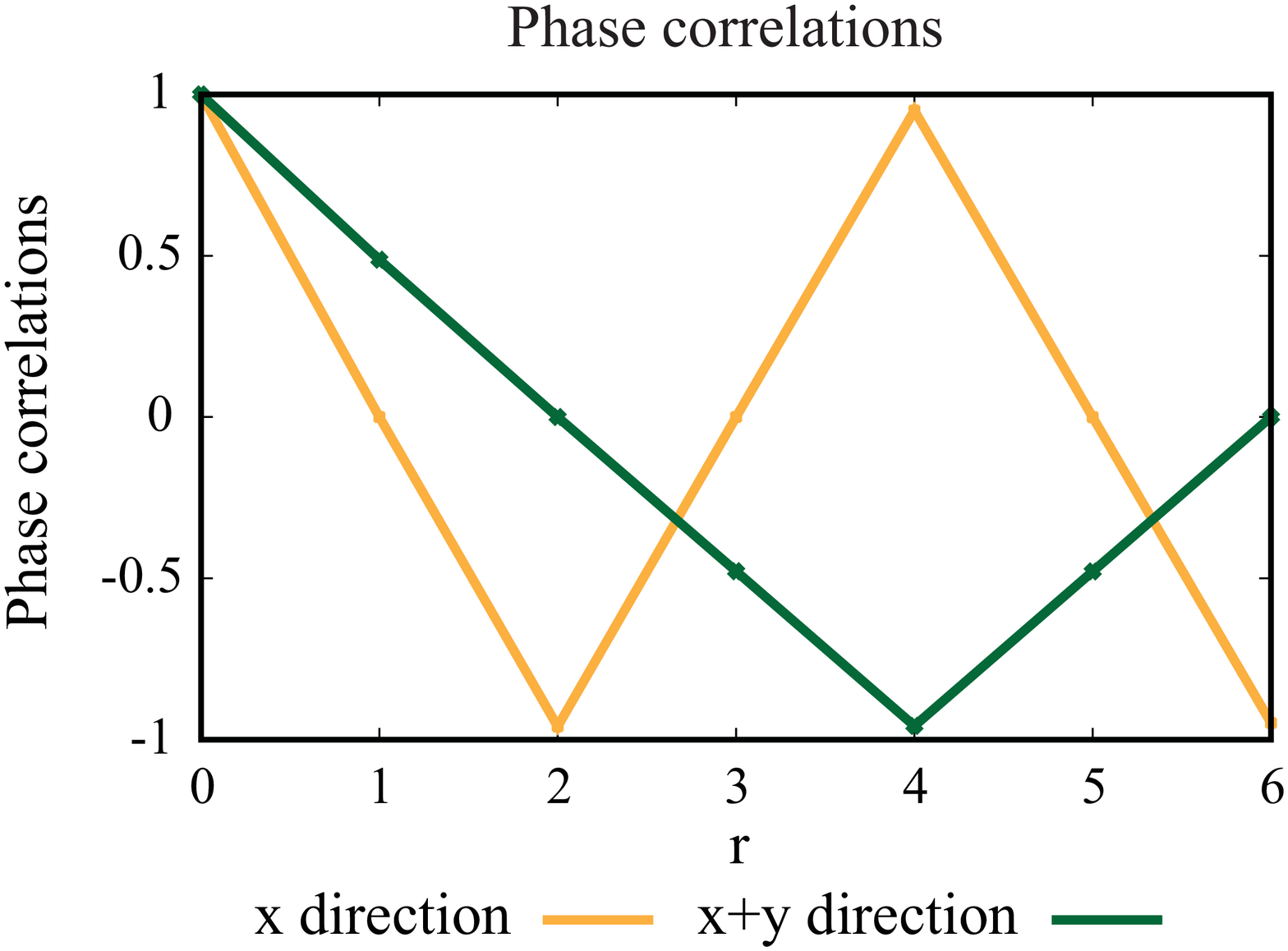}
\caption{(Color online) (Upper panel) 
Various order parameters that indicate SF+CSF phase.
$U_\tau=10, \ J_1/V=0.45, \ \mu/V=1.7$.
(Middle panel) Phase of boson operator.
(Lower panel) Correlation function $G_{\rm x}$ and $G_{\rm x+y}$ in the SF+CSF phase.
$U_\tau=10, \ J_1/V=0.45, \ J_2/V=0.24,\ \mu/V=1.7$.
System size $n_x=12$.
}
\label{SF+CSF2}
\end{figure}
\begin{figure}[h]
\centering
\includegraphics[width=8cm]{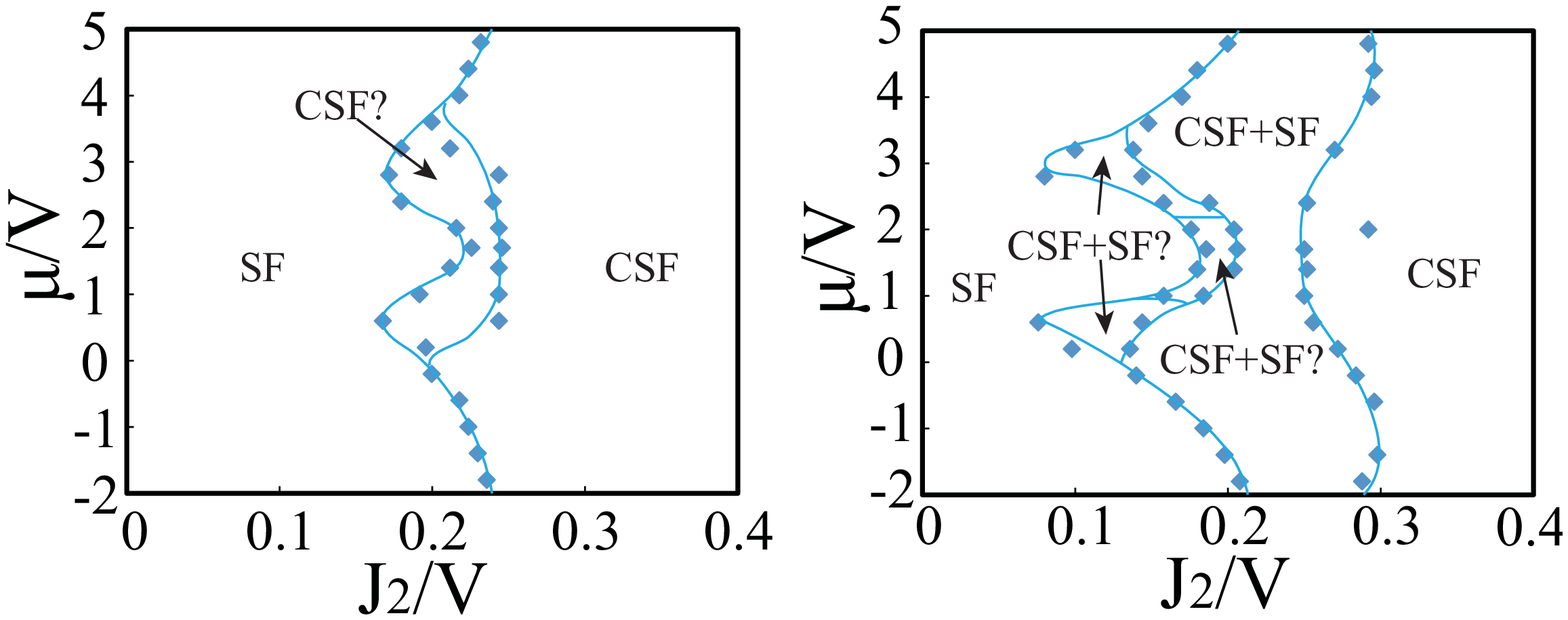}
\caption{(Color online) Phase diagram for $J_1/V=0.46, U_\tau=0.1$.
Case ${\cal O}$ (left panel) and Case ${\cal M}$ (right panel)
}
\label{PDJ1=23}
\end{figure}

In the Case ${\cal M}$, there appear additional states 
in which the SF and CSF coexist as the phase diagram in Fig.\ref{PDJ1=5}
shows..
The phase diagram itself is rather complicated and some of the phase boundaries
cannot be clarified by the present MC simulations.
However, we verified that the SF+CSF has a clear correlation in the phase
of the Bose condensate, which indicates a stable phase coherence.
This point will be discussed later on after showing the results for a larger $J_1$. 

We also show the phase diagram for $U_\tau=10$ in Fig.\ref{PDJ1=5_2}.
There is no substantial differences between the Cases ${\cal O}$ and ${\cal M}$.
However 
in contrast to the cases $U_\tau=0.1$, different configurations form in different
layers in the imaginary-time direction, i.e., there exists substantial quantum fluctuations
in the case $U_\tau=10$.
In particular in $\rho=1/2$ solid and the SL-CSFs, the sublattice, which is filled by
atoms, is different layer by layer.

Finally let us turn to the case with $J_1/V=0.45$.
In this case, the system exists in the SF for $J_2=0$ as shown in Fig.\ref{J20}.
We first show the phase diagrams for the Cases ${\cal O}$ and ${\cal M}$
with $U_\tau=10$ obtained by the MC simulation.
See Fig.\ref{PDJ1=22.5}.
Numerical studies show that the system has a rather stable phase diagram 
in the $(J_2/V$-$\mu/V)$-plain as in the case with $J_1/V=0.2$.
In the Case ${\cal O}$, the phase transition from the SF state to the CSF
takes place directly as $J_2$ is increased.
The specific heat $C$ exhibits a very sharp peak and the $L_{\rm NNN}$
and also $C_{\rm NNN}$ have a step-wise behavior at the phase transition point.
See Fig.\ref{SF+CSF2}.
Therefore we conclude that the phase transition from the SF to CSF is of first order
in this parameter region.
On the other hand for the Case ${\cal M}$, there exists the SF and CSF coexisting
phase between the genuine SF and the genuine CSF. 
We have reached this conclusion by the calculations of the order parameters,
which indicate that both the SF and CSF exist in that phase.
Calculation of the specific heat suggests that both of the two phase transitions,
SF$\leftrightarrow$SF+CSF and SF+CSF$\leftrightarrow$CSF, are of second order.
It is interesting to see a snapshot of the phase configuration in the SF+CSF phase.
See Fig.\ref{SF+CSF2}.
Similar results were obtained for the Case ${\cal M}$ with $J_1/V=0.1$.
The snapshot and the correlation functions indicate that there exists a stable
configuration in the SF+CSF.
In fact as the correlation functions $G_{\rm x}$ and $G_{\rm x+y}$ show, 
the boson phases rotate about $-{\pi \over 2}$ for 
one lattice spacing in the x-direction
and also about $-{\pi \over 2}$ for two lattice spacings in the (x+y)-direction
in that configurations.
From this observation, {\em the SF+CSF state breaks the $120^o$ rotational
symmetry and as a result it is threefold degenerate}.
(In the present calculation, the periodic boundary condition choses one out of the
three states.)
It is straightforward to calculate the energy of the above configuration
as the particle density is homogeneous.
In the Case ${\cal M}$, the energy per site per unit density is $(-4J_2-J_1)$.
On the other hand in the Case ${\cal O}$, for the A-sublattice $(-4J_2-J_1)$,
whereas for the B-sublattice $(4J_2-J_1)$.
Therefore average energy in the Case ${\cal O}$ is $-J_1$.
This means that the observed SF+CSF phase is stable in the parameter
region with an intermediate magnitude of $J_2$ in the Case ${\cal M}$, 
whereas not in the Case ${\cal O}$.
This explains the difference in the phase diagram of the two Cases. 

For the case $U_\tau=0.1$, the phase diagram has a rather complicated structure,
and clear identification of some phase boundaries is difficult as the
specific heat $C$ does not exhibit a consistent behavior for the measure in the
various directions in the parameter space.
Obtained phase diagram is shown Fig.\ref{PDJ1=23}, in which some of 
the phases are not definitive at present.
This fact indicates that the system has a complicated phase structure because of
the frustrations and some phase transitions are of strong first order.
More elaborated numerical methods than the local up-date MC simulation
is required to clarify the phase diagram.

\begin{figure}[h]
\centering
\includegraphics[width=4cm]{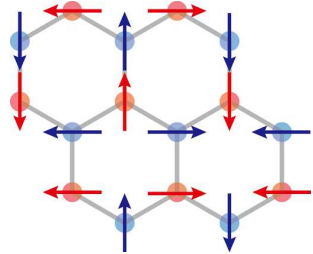}
\caption{(Color online) Phase configuration in the SF+CSF state observed in the snapshot
in Fig.\ref{SF+CSF2}.
}
\label{confgSF+CSF}
\end{figure}


\section{Discussion and Conclusion}

In this paper, we have introduced the extended HBHM that describes 
the dipolar hard-core bosonic gases captured in the honeycomb lattice.
In addition to  the NN hopping and on-site repulsion, there exist the 
NNN-complex hopping and the NN repulsion.
In order to study the case of strong-on-site repulsion, we employed
the slave-particle representation and used the coherent-state path
integral for the MC simulations.
To suppress the multi-particle states in the path integral, we added 
the $U'$-term to the action of the effective theory Eq.(\ref{partitionF}).
We considered two cases for the NNN-complex hopping, the one is that in 
the original Haldane model (Case ${\cal O}$) and the other is that in 
the so-called modified Haldane model (Case ${\cal M}$).
We showed that the different phase diagrams appear depending on the 
NNN hopping.

We first studied the simple case with the vanishing NNN hopping $J_2=0$.
There are four phases in the phase diagram including $\rho=1/2$-solid,
which has the DW properties.
In the weak quantum fluctuation case with $U_\tau=0.1$, the SS forms
in a small but finite parameter region in the $\rho=1/2$-solid.

Next we considered the weak NN hopping case, $J_1/V=0.1, \ U_\tau=0.1$
and $J_1/V=0.1, \ U_\tau=10.0$.
The phase diagram contains the CSF as well as the phase, which we call the SL-CSF.
In the SL-CSF, there exist the diagonal sublattice DW  order and 
the off-diagonal CSF order.
Therefore, it is a new kind of the SS.
We clarified the origin of the SL-CSF, i..e, the its existence stems from the
uncertainty relation between the particle number and phase of the boson operator.
Furthermore for the Case ${\cal M}$ with the small quantum fluctuation $U_\tau=0.1$,
the phase diagram contains the coexisting phase of the SF and CSF.
In this phase, the boson phase has a certain specific stable configuration as exhibited in Fig.\ref{confgSF+CSF}.
Finally we studied the case of the relatively large $J_1/V=0.45, \ 0.46$,
in which a large frustration exists.
For the case $U_\tau=10$, the stable phase diagrams are obtained for both the Case
${\cal O}$ and Case ${\cal M}$, whereas for $U_\tau=0.1$, 
the phase diagram is rather complicated and different phase boundaries are 
obtained by calculating the specific heat with increasing and decreasing the value
of the chemical potential $\mu$.
This indicates that there exist strong first-order phase transition line
in the phase diagram.
The phase diagram in Fig.\ref{PDJ1=23} was obtained by combining 
the calculations of the specific heat in both the directions.

In the present paper we considered the dipolar atoms by taking into account
the NN repulsions.
It is important to investigate how the nonlocal interaction of dipoles, which decays
$\sim 1/r^3$ with the distance $r$, influences the phase diagram.
We investigated  the effect of the NNN repulsion like 
$V'\sum_{\rm NNN}n_in_j$ for $V'=9.6$ and found that the instability
of $\rho=1/2$ solid in the phase diagram presented in Fig.\ref{J20} 
appears.
Detailed investigation of the effect of the nonlocal interactions is under study,
and we hope that results will be reported in a future publication.

As explained in Sec.III.B, the energy unit of this work is $(N_\tau k_{\rm B}T)$ with
$N_\tau=8$ in the present numerical study.
From the typical value of the magnetic dipole of atoms, we can estimate
the temperature $T$.
For $^{52}$Cr with $6\mu_{\rm B}$, where $\mu_{\rm B}$ is the Bohr magneton,
and the lattice spacing of the optical lattice $a$, $a=200$nm,
we have $V=3$nK by using $V=\mu_0(10\mu_{\rm B})^2/(4\pi a^3)$
where $\mu_0$ is the magnetic permeability of the vacuum.
Therefore the temperature of the system is estimated as $k_{\rm B}T=10^{-2}$nK.
Then the obtained phase diagrams can be regarded as the ground-state phase
diagrams.
It is interesting to study finite-temperature phase transitions of the SF and CSF.
In the present formalism, temperature of the system is controlled by
varying the value of $\Delta\tau$ as it is related to $T$, i.e., 
$k_{\rm B}T=1/(\Delta\tau N_\tau)$.
In the previous work on some related models, we investigated the finite-$T$
transitions by this methods \cite{RFIO}.
Rough estimation of the critical temperature of the SF, $T_c$, can be obtained by 
the following simple consideration.
We consider the case with $J_2=0$ for simplicity and decrease the value of
$\Delta\tau$.
From Eq.(\ref{partitionF}), the system is regarded as a quasi-two-dimensional
one, and the finite-$T$ phase transition of the SF is described by the O(2)
spin order-disorder phase transition.
The critical value of the phase transition $\Delta\tau_c$
is estimated as $J_{1}\Delta\tau_c\sim 1$ for $\rho=1/2$ (i.e., $X_{i,\ell}=\pi/4$).
From the phase diagram in Fig.\ref{J20}, the typical value of $J_1$ in {\em the SF
close to the $\rho=1/2$ Mott insulator} is $J_1/V=0.6$.
Therefore $1/\Delta\tau_c\sim 1.8$nK, which means $k_{\rm B}T_c \sim 0.2$nK.
We notice that this critical temperature is still low from the view point of the
present feasible experimental setup.
More detailed investigation of the finite-$T$ properties will be given in a future
publication.

It is interesting to study the present model with finite boundaries and 
investigate edge states.
In the previous paper \cite{Kuno} for the case $V=0$, we found that
certain specific state forms near the boundaries.
Recently some related work studied edge states in gapped state
of bosonic gases on the honeycomb lattice \cite{Guo}.

Another interesting system to be studied is the negative $J_1$ case of
the Hamiltonian in Eq.(\ref{HBHM}).
This system is closely related to the frustrated quantum spin system, 
and an interesting ground-state was proposed \cite{moat}.
It is rather straightforward to apply the extended MC in the present paper
to that system, and we hope that interesting results will be published in near future.

\bigskip

\acknowledgments
This work was partially supported by Grant-in-Aid
for Scientific Research from Japan Society for the 
Promotion of Science under Grant No.26400246.



\begin{thebibliography}{99}
\bibitem{note0}
To be more precise, the coherent state $|z\rangle$ with $|z|=1$ does not
strictly corresponds to the single-particle state, i.e., it contains small
but finite portions of other states. 
From this fact, the on-site $U'$-term in the action $S$ in Eq.(\ref{partitionF})
has been introduced to suppress the multi-particles states, 
although the numerical study shows that its effect on the phase diagram is negligibly small.
It should be remarked here that 
for the antiferromagnetic (AF) Heisenberg model, the present method derives a CP$^1$
field theory as an effective low-energy theory of the AF Heisenberg model.
It is known that the both systems belong to the same universality class \cite{AFHM}.
This fact indicates the reliability of the present method.
\bibitem{note1}
More precisely, the local constraint on the quantum fluctuations of the densities,
$\delta\rho_{bi}$ and $\delta\rho_{hi}$, $\delta\rho_{bi}+\delta\rho_{hi}=0$,
can be imposed without any difficulty by using a Lagrange multiplier.
This manipulation results in appearance of a gauge field in the imaginary-time
derivative term of $\theta_{bi}$ and $\theta_{hi}$.
However, this modification does not substantially influence on the phase diagrams.
\end{thebibliography}
\end{document}